 \documentclass[twocolumn,amsmath,amssymb,prb]{revtex4}
\usepackage{graphicx}

\newcommand{ \Sec }[1]{Sec.~\ref{sec:#1}}
\newcommand{ \Appendix }[1]{Appendix \ref{sec:#1}}

\newcommand{ \Eq  }[1]{Eq.~(\ref{#1})}
\newcommand{ \Eqs }[2]{Eqs.~(\ref{#1}) and (\ref{#2})}

\newcommand{ \Table }[1]{Table \ref{tab:#1}}

\newcommand{ \Ref  }[1]{Ref.~\onlinecite{#1}}
\newcommand{ \Refs }[2]{Refs.~\onlinecite{#1} and \onlinecite{#2}}

\newcommand{ \Fig     }[1]{Fig.~\ref{fig:#1}}

\newcommand{ \Figure  }[1]{Figure \ref{fig:#1}}

\newcommand{ \etal }{\textit{et al.}}

\newcommand{ \apriori     }{\textit{a priori}}

\newcommand{ \BE }{ \begin{equation} }
\newcommand{ \EE }{ \end{equation} }

\newcommand{ \BEA }{ \begin{eqnarray} }
\newcommand{ \EEA }{ \end{eqnarray} }

\newcommand{ \BSUB }{\begin{subequations}}
\newcommand{ \ESUB }{\end{subequations}}

\newcommand{ \BFIG }{\begin{figure}}
\newcommand{ \EFIG }{\end{figure}}

\newcommand{ \BW }{ \begin{widetext} }
\newcommand{ \EW }{ \end{widetext} }

\newcommand{ \del  }{ \partial }
\newcommand{ \half }{ \frac{1}{2} }

\newcommand{ \Schroedinger }{Schr\"{o}dinger~}

\newcommand{ \QM }{ \mathrm{QM} }
\newcommand{ \MM }{ \mathrm{MM} }
\newcommand{ \QMMM }{ \mathrm{QM/MM} }

\newcommand{ \SC }{ \mathrm{sc} }
\newcommand{ \MF }{ \mathrm{MF} }

\newcommand{ \vdw }{ \mathrm{vdw} }

\newcommand{ \HF }{ \mathrm{HF} }
\newcommand{ \MP }{ \mathrm{MP2} }

\newcommand{ \AQMMM }{ A }
\newcommand{ \AMF }{ A^\mathrm{MF} }
\newcommand{ \Avp }{ \tilde{A} }

\newcommand{ \EMF }{ E^\mathrm{MF} }

\newcommand{ \AMM }{ \Delta A_\MM }

\newcommand{ \Afluc }{ \Delta A_\mathrm{fluc} }
\newcommand{ \Efluc }{ \Delta E }
\newcommand{ \EeffQMfluc }{ \Delta E_\mathrm{fluc} }

\newcommand{ \ATS }{ \Delta A^\ddagger }
\newcommand{ \Areact }{ \Delta A_\mathrm{r} }

\newcommand{ \GTS }{ \Delta G^\ddagger }
\newcommand{ \Greact }{ \Delta G_\mathrm{r} }

\newcommand{ \HQM }{ \hat{H}_\QM }
\newcommand{ \EQM }{ E_\QM }
\newcommand{ \EMM }{ E_\MM }

\newcommand{ \Eeff }{ \mathcal{E} }
\newcommand{ \EeffQM }{ \Eeff_\QM }

\newcommand{ \EeffMM }{ \Eeff_\MM }

\newcommand{ \EeffMMvdw }{ \mathcal{E}_\MM }

\newcommand{ \Rvec }{ \mathbf{R} }
\newcommand{ \Rvecopt }{ \mathbf{R}^{*} }

 \newcommand{ \RMM }{ \mathbf{R}^{+} }

\newcommand{ \intMM }{ - \frac{1}{\beta} \ln \int d\RMM }

\newcommand{ \vmm }{ v_\MM }

\newcommand{ \vmmref }{ v_\MM^\circ }

\newcommand{ \Phivec }{ \mathbf{v} }

\newcommand{ \PhivecMM }{ \Phivec_\MM }
\newcommand{ \PhivecSC }{ \Phivec^\SC }
\newcommand{ \Phivecref }{ \Phivec^\circ }
\newcommand{ \Phivecave }{ \Phivec }

\newcommand{ \PhivecSCHF }{ \Phivec^\SC_\mathrm{HF} }
\newcommand{ \PhivecSCMP }{ \Phivec^\SC_\mathrm{MP2} }

\newcommand{ \Qvecref }{ \mathbf{Q}^\circ }
\newcommand{ \Qvec }{ \mathbf{Q} }
\newcommand{ \QvecSC }{ \mathbf{Q}^\SC }
\newcommand{ \Qop }{ \hat{Q} }
\newcommand{ \Qopvec }{ \hat{\Qvec} }
\newcommand{ \Qvecop }{ \hat{\Qvec} }

\newcommand{ \QvecSCHF }{ \mathbf{Q}^\SC_\mathrm{HF} }
\newcommand{ \QvecSCMP }{ \mathbf{Q}^\SC_\mathrm{MP2} }

\newcommand{ \PsiSC }{ \Psi^\SC }
\newcommand{ \lambdaSC }{ \lambda^\SC }

\newcommand{ \xvec }{ \mathbf{x} }
\newcommand{ \Xvec }{ \mathbf{X} }
\newcommand{ \rvec }{ \mathbf{r} }

\newcommand{ \yvec }{ \mathbf{y} }

\newcommand{ \Psiref }{ \Psi^\circ }

\newcommand{ \chiQM   }{ \boldsymbol{\chi}_\QM }
\newcommand{ \sigmaMM }{ \boldsymbol{\sigma}_\MM }

\newcommand{ \rhoop }{ \hat{\rho} }
\newcommand{ \rhoSC }{ \rho^\SC }

\newcommand{ \esp   }{ v }
\newcommand{ \espMM }{ \esp_\MM }
\newcommand{ \espSC }{ \esp^\SC }

\newcommand{ \Asolv }{ \Delta A_\mathrm{solv} }

\newcommand{ \Psiave }{ \tilde{\Psi} }

\newcommand{ \SNtwo }{S$_\mathrm{N}$2~}

\newcommand{ \Ropt }{ \Rvec^{*} }
\newcommand{ \RC   }{ s }

\newcommand{ \AMMgrad }{ \mathbf{G} }

\newcommand{ \rCNTS }{ r^\ddagger(\mathrm{C-N}) }
\newcommand{ \rCClTS }{ r^\ddagger(\mathrm{C-Cl}) }

\newcommand{ \EQMtilde }{ \tilde{E}_\QM }

\newcommand{ \Qtot }{ Q_{\mathrm{tot}} }

\newcommand{ \Gvec }{ \mathbf{G} }

\newcommand{ \rqm  }{ \mathbf{r}_\QM }
\newcommand{ \rqmi }{ \mathbf{r}_{\QM,i} }
\newcommand{ \rqmj }{ \mathbf{r}_{\QM,j} }

\newcommand{ \rmm  }{ \mathbf{r}_\MM }

\newcommand{ \rmmref }{ \rmm^\circ }

\newcommand{ \Heff    }{ \hat{H}_\mathrm{eff} }
\newcommand{ \Heffref }{ \Heff^\circ }
\newcommand{ \Hqm     }{ \hat{H}_\QM }

\newcommand{ \MMave }{ \frac{ 1 }{ L } \sum_{ \tau = 1 }^{ L } }

\newcommand{ \sumqmi }{ \sum_i^\QM }
\newcommand{ \sumqmj }{ \sum_j^\QM }

\newcommand{ \Eeffref }{ \langle \Psi | \Heff | \Psi \rangle^\circ }
\newcommand{ \Eref }{ E_\mathrm{ref} }

\newcommand{ \Qref }{ Q^\circ }

\newcommand{ \Einternal }{ E_1^\circ }

\newcommand{ \rderiv }[1]{ \frac{ \del #1 }{ \del \rqm } }
\newcommand{ \rderivTot }{ \frac{ \del }{ \del \rqm } }

\newcommand{ \RefYang }{\Ref{Yang_QMMM_MFEP08}}
\newcommand{ \citeYang }{\cite{Yang_QMMM_MFEP08}}

\newcommand{ \Angyan }{\'{A}ngy\'{a}n}

\newcommand{ \VintMF }{ \hat{V}_\mathrm{QM/MM}^\MF }

\newcommand{ \siteA }{ a }
\newcommand{ \siteB }{ b }

\begin{document}

\title{Variational and perturbative formulations of
QM/MM free energy with mean-field embedding and its analytical gradients}

\author{Takeshi Yamamoto}
\affiliation{Department of Chemistry, Kyoto University, Kyoto 606-8502, Japan}
\email{yamamoto@kuchem.kyoto-u.ac.jp}

\begin{abstract}
Conventional quantum chemical solvation theories are based on the
mean-field embedding approximation. That is,
the electronic wavefunction is calculated in the presence of
the mean field of the environment.
In this paper a direct quantum mechanical/molecular mechanical
(QM/MM) analog of such a mean-field theory
is formulated based on variational and perturbative frameworks.
In the variational framework, an appropriate QM/MM free energy functional
is defined and is minimized in terms of the trial wavefunction that
best approximates the true QM wavefunction in a statistically
averaged sense. Analytical free energy gradient
is obtained, which takes the form of the gradient
of effective QM energy calculated in the averaged MM potential.
In the perturbative framework, the above variational
procedure is shown to be equivalent with the first-order
expansion of the QM energy (in the exact free energy expression)
about the self-consistent reference field.
This helps understand the relation between the variational procedure
and the exact QM/MM free energy as well as existing QM/MM theories.
Based on this, several ways are discussed
for evaluating non-mean-field effects
(i.e., statistical fluctuations of the QM wavefunction)
that are neglected in the mean-field calculation.
As an illustration, the method is applied to
an \SNtwo Menshutkin reaction in water,
$
    \mathrm{
      NH_3 + CH_3 Cl
      \rightarrow
      NH_3CH_3^{+} + Cl^{-}
    }
    ,
$
for which free energy profiles are obtained at the HF, MP2, B3LYP,
and BH\&HLYP levels by integrating the free energy gradient.
Non-mean-field effects are evaluated
to be $< 0.5$ kcal/mol using a
Gaussian fluctuation model for the environment,
which suggests that those effects are rather
small for the present reaction in water.
\end{abstract}

\maketitle

%-----------------------------------------------------------------------------
\section{\label{sec:intro}Introduction}

A combined quantum mechanical/molecular mechanical (QM/MM)
method is a powerful computational tool for studying chemical reactions
in solution and in biological systems.\cite{Warshel_Book,Cramer_Book}
It treats a chemically active part
of the entire system with accurate QM methods while the rest of the system
with MM force fields. The quality of a given QM/MM calculation depends primarily
on the electronic structure method used. In the calculation of
statistical properties like free energy, it is also important to
adequately sample the relevant phase space.\cite{Warshel_QMMM_Pitfall05}
However, this phase space sampling
is very demanding computationally, because one needs to
calculate QM electronic energy for a large number of
statistical samples.
One can ensure sufficient statistics by using fast semiempirical methods,
but the resulting energetics may be less satisfactory than obtained with
ab initio methods. On the other hand, highly correlated QM methods
require too much computational time
and thus it becomes difficult to explore the phase space.

%...................................................................

A variety of approaches have been proposed
in order to address the above trade-off between accuracy and efficiency.
One approach is a family of dual-level methods, in which a
classical or semiempirical potential is used for statistical sampling
and an accurate QM method for energetic corrections.\cite{Warshel_QMMM_DualLevel95,Warshel_QMMM_DualLevel98,Warshel_QMMM_DualLevel02,Warshel_QMMM_DualLevel06,Wood_QMMM_DualLevel99,Wood_QMMM_DualLevel00,Wood_QMMM_DualLevel02,Ryde_QMMM_DualLevel_PRL,Ryde_QMMM_DualLevel,Moliner_QMMM_IC_Mensh,Dupuis_QMMM_DualLevel,Roitberg_QMMM_Chorismate05}
Another approach is to introduce some approximation to the QM--MM electrostatic
interactions in order to reduce the number of QM calculations.
Our main interest in this paper is in the second approach above.
In particular, we are concerned with the following
three \textit{embedding} schemes that
prescribe how to calculate the QM wavefunction in the MM environment:

%...................................................................

(1) \textit{Gas-phase embedding scheme.}
This scheme totally neglects electrostatic perturbations of
the MM environment on the QM subsystem.
The QM wavefunction is calculated \apriori{} in the gas phase, and
the resulting charge density or partial charges are embedded
into the MM environment.
The reaction path is also determined by the gas-phase calculation.
The free energy profile in solution is obtained
via free energy perturbation (FEP) calculations along the
pre-determined reaction path. This approach was first utilized by
Jorgensen and co-workers\cite{Jorgensen_QMFE_SN2,Jorgensen_QMFE_Review,Jorgensen_QMFE_DielsAlder,Jorgensen_QMFE_Claisen} to study organic reactions in solution,
and later by Kollman and co-workers\cite{Kollman_QMFE_Amide,Kollman_QMFE_Catechol,Kollman_QMFE_review}
to study enzyme reactions.
It should be noted however that this approach may not be appropriate
for a certain class of enzyme reactions.\cite{Yang_QMMM_pseudobond}

%...................................................................

(2) \textit{Mean-field embedding scheme.}
This method calculates the QM wavefunction in the presence
of the mean field of the environment.
The averaged polarization (or distortion) of the QM wavefunction is thus
correctly taken into account,
while statistical fluctuations of the QM wavefunction
are totally neglected.
Indeed, this mean-field approximation has been the basis of many
conventional solvation models
like the PCM\cite{Tomasi_Review1,Tomasi_Review2}
and RISM-SCF\cite{Tenno_RISM,Sato_RISM,Sato_3DRISM,Hirata_Book} methods.
The mean-field idea has also been applied to the QM/MM framework
by several authors.
For example, Aguilar and co-workers
\cite{Aguilar_ASEP_Program,Aguilar_ASEP_Opt,Aguilar_ASEP_Mensh,Aguilar_ASEP_Stark,Aguilar_ASEP_Compare}
performed
geometry optimization on an approximate QM/MM free energy surface using
the averaged solvent electrostatic potential (ASEP)/MD method.
More recently, the mean-field idea was exploited
by Warshel and co-workers\cite{Warshel_QMMM_Meanfield08}
in order to accelerate QM/MM calculation of
solvation free energy.

%...................................................................

(3) \textit{Polarizable embedding scheme.}
This method first develops a polarizable model of the QM subsystem and then
embeds the resulting model into the MM environment. The polarizable QM model can be developed,
for example, by Taylor expanding the QM electronic energy up to second order.
\cite{Morita_CRK1,Morita_CRK2,Naka_RISM_Fluc,Yang_QMMM_RPP,Yang_QMMM_MFEP07,Yang_QMMM_MFEP08,Yang_QMMM_Review,Higashi_MCMM}
The QM/MM minimum free energy path (MFEP) method by
Yang and co-workers
\cite{Yang_QMMM_RPP,Yang_QMMM_MFEP07,Yang_QMMM_MFEP08,Yang_QMMM_Review}
is based on this perturbative expansion idea, which has been applied
to chemical reactions in solution and enzymes.
Among the three embedding schemes above,
the polarizable one is most accurate by allowing statistical
fluctuations of the QM wavefunction.

%...................................................................

The first goal of this paper is to formulate the mean-field embedding scheme
above by starting from
a variational principle for the QM/MM free energy (\Sec{MF_var}).
As mentioned above, 
conventional solvation models are based on the mean-field embedding
approximation. They often start with a variational principle for the following
free energy
\cite{Tomasi_Review1,Tomasi_Review2,Angyan_Review}
\BE
\label{A_conv}
    A( \rqm )
    =
    \langle \Psi | \HQM | \Psi \rangle
    +
    \Asolv[ \Psi ]
    .
\EE
Minimization of $A( \rqm )$ in terms of $\Psi$
gives a nonlinear \Schroedinger equation for $\Psi$
that is subject to the mean field of the environment.
Very often, analytical gradient of free energy,
$\del A( \rqm ) / \del \rqm$, is obtained by utilizing
the variational nature of $A( \rqm )$.
Since those solvation models are quite successful
in studying solution-phase chemistry,
it is natural to try to extend them to the QM/MM framework.
The main benefits of this extension are as follows.
First, QM/MM models can describe
inhomogeneous as well as homogeneous environments
on an equal theoretical footing.
This makes it more straightforward to compare the chemical reactivity
of a system in different environments (e.g., in solution and enzymes).
Second, since the mean-field QM wavefunction is calculated
only for a ``batch'' of MM configurations, the number of
QM calculations can be made significantly smaller than
a direct QM/MM statistical calculation.
As mentioned above, such a mean-field QM/MM approach has been explored
by several authors in the literature.
For example, \Angyan{}\cite{Angyan_Review} discussed
such a method quite a few years ago based on a variational principle and
linear-response approximation (LRA).
More recently, Kato and co-workers\cite{Higashi_LRFE,Higashi_LRFE2}
developed the QM/MM LRFE method using a different type of
variational/LRA idea and applied it
to chemical reactions in solution and enzymes.
On the other hand, Aguilar and co-workers\cite{Aguilar_ASEP_Opt}
took a different approach in the ASEP/MD method,
where they did not invoke a variational principle nor LRA but
rather approximated the free energy gradient as follows:
\BE
\label{Aguilar_grad}
    \rderiv{ A( \rqm ) }
    =
    \left\langle \rderiv{ E( \rqm, \rmm ) } \right\rangle
    \simeq
    \rderivTot \langle E( \rqm, \rmm ) \rangle
    .
\EE
Here, $E( \rqm, \rmm )$ is the total energy of the QM/MM system
and $\langle \cdots \rangle$ denotes
the statistical average over MM degrees of freedom.
Note that in \Eq{Aguilar_grad}
``the average of the energy gradient'' 
(the exact expression) is replaced
by ``the gradient of the averaged energy''
in the spirit of the mean-field approximation.
While Aguilar \etal{} demonstrated its accuracy
via comparison with direct QM/MM calculations,\cite{Aguilar_ASEP_Opt}
the detailed derivation of \Eq{Aguilar_grad} was not provided
and it was used as an ansatz.
Therefore, our first aim in this paper is
(i) to formulate a mean-field
QM/MM framework by starting from a variational principle
(but not invoking the LRA), (ii) obtain analytical gradient of
the associated free energy, and (iii) discuss a possible rationale
for the approximate gradient in \Eq{Aguilar_grad}.

%...................................................................

The second goal of this paper is to understand
the relation of the above variational/mean-field procedure
with the underlying exact QM/MM free energy
as well as existing QM/MM theories (Secs.~\ref{sec:MF_pert} and \ref{sec:fluc}).
First, it is shown that
the above variational procedure is equivalent with
the first-order expansion of effective QM energy
(in the exact free energy expression)
about the self-consistent reference field.
As mentioned above, the QM/MM-MFEP method
\cite{Yang_QMMM_MFEP07,Yang_QMMM_MFEP08}
is based on this type of perturbative expansions.
Therefore, it is interesting to compare the present approach
with the QM/MM-MFEP method in detail (\Appendix{QMMM_MFEP}).
From this comparison it follows that
the variational procedure is essentially equivalent with
Model 3 of the QM/MM-MFEP method
with charge response kernel $\chi$ neglected.
Note however that the full version of Model 3 includes that response kernel
$\chi$ and thus it is more accurate
by describing statistical fluctuations of the QM wavefunction.
Therefore, in \Sec{fluc}
we also discuss several possible ways for evaluating such non-mean-field effects
on top of the variational/mean-field calculation.

%...................................................................

As an illustration, the present method is applied to
an \SNtwo reaction in water (\Sec{app}).
Free energy profiles are obtained by integrating the free energy
gradient and they are compared
with free energy perturbation (FEP) results.
Non-mean-field effects are also evaluated using a Gaussian
fluctuation model for the environment. The obtained results suggest
that the non-mean-field effects are rather
small for the present reaction in water.

% \clearpage

%=============================================================================
\section{\label{sec:theory}Methodology}

%=============================================================================
\subsection{\label{sec:qmmm_fe}The underlying QM/MM free energy}

We consider the following QM/MM free energy
(or the potential of mean forces acting on QM atoms)
\BE
\label{AQMMM}
    \AQMMM( \Rvec )
	=
	- \frac{ 1 }{ \beta }
	\ln
	\int d\RMM
	\exp
	\{
	  - \beta
      E( \Rvec, \RMM )
	\}
    ,
\EE
where $\Rvec$ and $\RMM$ are Cartesian coordinates of the QM and MM atoms,
respectively, $\beta = 1/k_B T$ is the reciprocal temperature, and
$E( \Rvec, \RMM )$ is the total energy given by
\BE
\label{total_energy}
    E( \Rvec, \RMM )
    =
    \EeffQM( \Rvec, \PhivecMM( \Rvec, \RMM ) )
    +
    \EeffMMvdw( \Rvec, \RMM )
    .
\EE
Here, $\EeffQM$ is the electronic energy of the QM subsystem
in the presence of an external electrostatic field
(called the effective QM energy).
In the standard electronic embedding scheme, it is defined via the
following \Schroedinger equation
\BE
\label{SE_cont}
    [
      \HQM
	  +
	  \int d\xvec
      \rhoop( \xvec )
	  \esp'( \xvec )
	]
	| \Psi[ \Rvec, \esp' ] \rangle
	=
	\EeffQM[ \Rvec, \esp' ]
	| \Psi[ \Rvec, \esp' ] \rangle
\EE
(note that the prime symbol will be attached
on variables and functions of ``dummy'' nature).
$\HQM$ is the QM Hamiltonian in the gas phase,
$\rhoop( \xvec )$ is the QM charge density operator,
\BE
\label{rho}
    \rhoop( \xvec )
	=
	\sum^\mathrm{nuc}_\siteA
	Z_\siteA
	\delta( \xvec - \Rvec_\siteA )
	-
	\sum^\mathrm{ele}_i
	\delta( \xvec - \hat{\rvec}_i )
    ,
\EE
and $\esp'( \xvec )$ is an (arbitrary) external electrostatic field.
In this paper we will utilize a discretized approximation to
\Eq{SE_cont} given by
\BE
\label{SE}
    [ \hat{H}_\QM + \sum_\siteA \hat{Q}_\siteA v'_\siteA ]
	| \Psi( \Rvec, \Phivec' ) \rangle
	=
	\Eeff_\QM( \Rvec, \Phivec' )
	| \Psi( \Rvec, \Phivec' ) \rangle
    ,
\EE
where $\{ \Qop_\siteA \}$ are a set of ``partial charge'' operators
associated with QM atoms $\{ \Rvec_\siteA \}$, and
$v'_\siteA = v'( \Rvec_\siteA )$.
The motivation for using \Eq{SE} is that
the external field can be parametrized by
an $N$-dimensional vector $\Phivec' = ( v'_1, \ldots, v'_N )$,
where $N$ is the number of QM atoms.
This fact makes the following discussion somewhat simpler.
Nevertheless, we stress that there is no fundamental difficulty in using
the original \Schroedinger equation in \Eq{SE_cont};
see \Appendix{cont_rep} for such a formulation.
In \Appendix{ESP_charge}, we summarize the present
definition of the partial charge operator $\Qvecop = (\Qop_1,\ldots,\Qop_N)$
based on the electrostatic potential (ESP) fitting procedure.\cite{RESP}

%...................................................................

Now going back to \Eq{total_energy},
$\PhivecMM( \Rvec, \RMM ) = ( v_{\MM,1}, \ldots, v_{ \MM, N } )$
are MM electrostatic potentials acting on QM atoms,
\BE
\label{vmm_alpha}
    v_{ \MM, \siteA }
    =
    \vmm( \Rvec_\siteA, \RMM )
    ,
\EE
where
\BE
\label{vmm}
    \vmm( \xvec, \RMM )
    =
	\sum_l
	\frac
		{ q_l^\MM }
		{ | \xvec - \RMM_l | }
\EE
with $\{ q^\MM_l \}$ being partial charges of the MM atoms.
$\EeffMMvdw( \Rvec, \RMM )$ is
the sum of the van der Waals interactions between QM--MM subsystems
and the internal energy of the MM subsystem,
\BE
    \EeffMMvdw( \Rvec, \RMM )
	=
	E_{\QMMM}^\vdw( \Rvec, \RMM )
	+
	\EMM( \RMM )
	.
\EE
In the following we will sometimes drop
the arguments of $\PhivecMM$ and $\EeffMM$
for notational simplicity, i.e.
$\PhivecMM = \PhivecMM( \Rvec, \RMM )$ and 
$\EeffMM = \EeffMM( \Rvec, \RMM )$.

%-----------------------------------------------------------------------------

\subsection{\label{sec:MF_var}Variational approach for mean-field embedding}

The free energy in \Eq{AQMMM} may be rewritten as
\BW
\BE
\label{AQMMM_explicit}
    \AQMMM( \Rvec )
	=
	- \frac{ 1 }{ \beta }
	\ln
	\int d\RMM
	\exp
	\{
	  - \beta
	  [
		\langle \Psi( \Rvec, \PhivecMM ) |
		\HQM
	   + \Qvecop \cdot \PhivecMM
	   | \Psi( \Rvec, \PhivecMM ) \rangle
		+
		\EeffMMvdw
	  ]
	\}
    ,
\EE
\EW
by explicitly writing $\EeffQM( \Rvec, \PhivecMM )$
in terms of the QM wavefunction.
Note that $\PhivecMM$ always stands for $\PhivecMM( \Rvec, \RMM )$
as mentioned above.
A direct evaluation of $\AQMMM( \Rvec )$ is computationally demanding
because $\Psi( \Rvec, \PhivecMM )$ depends on $\RMM$ through
$\PhivecMM = \PhivecMM( \Rvec, \RMM )$.
To avoid repeated QM calculations,
let us replace the true wavefunction
$\Psi( \Rvec, \PhivecMM )$ by some trial one $\Psiave( \Rvec )$
that best approximates the true wavefunction in a statistically
averaged sense. To do so, we consider a free energy functional of the form
\BW
\BE
\label{A_vp}
    \Avp[ \Rvec, \Psiave ]
	=
    \intMM
	\exp
	\{
	  - \beta
	  [
		\langle
		\Psiave
		|
		\HQM
		+
		\Qvecop \cdot \PhivecMM
		|
		\Psiave
		\rangle
		+
		\EeffMMvdw
	  ]
	\}
    .
\EE
Since the following inequality holds by definition for arbitrary
$\PhivecMM = \PhivecMM( \Rvec, \RMM )$ [we assume
that $\Psi(\Rvec,\PhivecMM )$ is
the ground state of $\HQM + \Qvecop \cdot \PhivecMM$],
\BE
	\langle
      \Psi( \Rvec, \PhivecMM )
      |
      \HQM
      +
      \Qvecop \cdot \PhivecMM
      |
      \Psi( \Rvec, \PhivecMM )
	\rangle
    \;
    \leqslant
    \;
   	\langle
	  \Psiave( \Rvec )
      |
      \HQM
      +
      \Qvecop \cdot \PhivecMM
      |
      \Psiave( \Rvec )
	\rangle
    ,
\EE
\EW
we obtain a variational principle for free energy
\BE
\label{vp}
    \AQMMM( \Rvec )
    \leqslant
    \Avp[ \Rvec, \Psiave ]
    .
\EE
Namely, $\Avp[ \Rvec, \Psiave ]$ is a strict upper bound on $\AQMMM( \Rvec )$,
and the best approximation to $\AQMMM( \Rvec )$ is obtained by
minimizing $\Avp[ \Rvec, \Psiave ]$ with respect to $\Psiave$.
This variational principle is indeed a direct QM/MM analog
of the standard ones used in conventional solvation
theories.\cite{Angyan_Review,Tomasi_Review1,Tomasi_Review2}
By minimizing the following Lagrangian to account for
the normalization of $\Psiave$,
\BE
    L[ \Rvec, \Psiave, \lambda ]
	=
	\Avp[ \Rvec, \Psiave ]
    -
    \lambda \{ \langle \Psiave | \Psiave \rangle - 1 \}
    ,
\EE
we obtain the following stationary condition for $\Psiave$:
\BE
\label{NLSE}
    \left[
	  \HQM + \Qvecop \cdot \ll \PhivecMM \gg_{ \Qvec[ \Psiave ] }
	\right]
	| \Psiave \rangle
	=
	\lambda
	| \Psiave \rangle
    .
\EE
Here $\ll \cdots \gg$ represents the statistical average over
MM degrees of freedom,
\BE
\label{MM_ave}
    \ll \cdots \gg_{ \Qvec' }
	=
	\frac
	{
	  \int d\RMM
      e^{ - \beta [ \Qvec' \cdot \PhivecMM
          + \EeffMMvdw ] }
      ( \cdots )
	}
	{
	  \int d\RMM
      e^{ - \beta [ \Qvec' \cdot \PhivecMM
          + \EeffMMvdw ] }
	}
    ,
\EE
and $\Qvec[ \Psi' ] = \langle \Psi' | \Qvecop | \Psi' \rangle$.
[In this paper the double bracket $\ll \cdots \gg$ indicates
that the average is of ``classical'' nature, i.e., it does not require
repeated QM calculations.]
Since \Eq{NLSE} is nonlinear with respect to $\Psiave$,
it is usually solved via iteration.
It follows from comparison between \Eqs{SE}{NLSE}
that $\Psiave$ and $\lambda$ may be written as
\BSUB
\BEA
    \Psiave & = & \Psi( \Rvec, \PhivecSC )
    ,
\\
    \lambda & = & \EeffQM( \Rvec, \PhivecSC )
    ,
\EEA
\ESUB
where $\PhivecSC$ is the self-consistent response field determined by
\BSUB
\label{SC_cond}
\BEA
    \PhivecSC( \Rvec )
    & = &
    \ll \PhivecMM( \Rvec, \RMM ) \gg_{ \QvecSC }
    ,
	\\
    \QvecSC( \Rvec )
	& = &
    \langle \PsiSC | \Qvecop | \PsiSC \rangle
    ,
\EEA
\ESUB
with $\PsiSC = \Psi( \Rvec, \PhivecSC )$.
In this paper, \Eq{SC_cond} will be called
the self-consistent (embedding) condition.
The minimum value of the free energy functional is obtained by
inserting $\Psiave = \PsiSC$ into \Eq{A_vp}:
\BW
\BEA
    \min_{ \Psiave } \Avp[ \Rvec, \Psiave ]
	& = &
    \intMM
	\exp
	\{
	  - \beta
	  [
		\langle
		\PsiSC
		|
		\HQM
		+
		\Qvecop \cdot \PhivecMM
		|
		\PsiSC
		\rangle
		+
		\EeffMMvdw
	  ]
	\}
	\nonumber
	\\
	& = &
	\langle \PsiSC | \HQM | \PsiSC \rangle
	+
	\AMM( \Rvec, \QvecSC )
    \nonumber
	\\
	& \equiv &
    \AMF( \Rvec )
    ,
\label{AMF_var}
\EEA
\EW
where $\AMM$ is defined by
\BE
\label{AMM}
    \AMM( \Rvec, \Qvec' )
	=
	\intMM
	e
    ^{
      - \beta
	  [ \Qvec' \cdot \PhivecMM + \EeffMMvdw ]
	}
    .
\EE
Note that $\langle \PsiSC | \HQM | \PsiSC \rangle$
has been extracted from the integral over $\RMM$
since it is independent of $\RMM$.
By the last line of \Eq{AMF_var}, we define the QM/MM free energy
with mean-field embedding approximation, $\AMF( \Rvec )$.

%...................................................................

The analytical gradient of $\AMF( \Rvec )$ can be obtained using a
standard procedure as follows.
First, we rewrite the $\AMF( \Rvec )$ in terms of the Lagrangian,
\BE
    \AMF( \Rvec )
	=
	\Avp[ \Rvec, \PsiSC ]
	=
	L[ \Rvec, \PsiSC, \lambdaSC ]
\EE
with $\lambdaSC = \EeffQM( \Rvec, \PhivecSC )$.
Recalling that $L[ \Rvec, \Psiave, \lambda ]$ is stationary with respect to
$\Psiave$ and $\lambda$, we obtain
\BE
    \frac{ \del }{ \del \Rvec }
	\AMF( \Rvec )
	=
	\left.
	\frac
		{ \del L[ \Rvec, \Psiave, \lambda ] }
		{ \del \Rvec }
	\right|
	_{ \Psiave = \PsiSC, \lambda = \lambdaSC }
    ,
\EE
where the $\Rvec$ derivative in the right-hand side
does not act on $\Psiave$ nor $\lambda$.
We then obtain the analytical gradient in \Eq{grad},
which will be discussed in the next section.

%-----------------------------------------------------------------------------

\subsection{\label{sec:MF_pert}Perturbative approach for mean-field embedding}

$\AMF( \Rvec )$ in \Eq{AMF_var} can also be obtained
from the exact $A( \Rvec )$ in \Eq{AQMMM}
by Taylor expanding the effective QM energy
$\EeffQM( \Rvec, \PhivecMM )$ up to first order:
\begin{multline}
\label{1st_order_expansion}
    \EeffQM( \Rvec, \PhivecMM )
	\simeq
	\EeffQM( \Rvec, \Phivecref )
\\
	+
	\left.
	  \frac{ \del \EeffQM( \Rvec, \Phivec' ) }
		   { \del \Phivec' }
	\right|
	_{ \Phivec' = \Phivecref }
	\cdot
	( \PhivecMM - \Phivecref )
    .
\end{multline}
Here, $\Phivecref = \Phivecref( \Rvec )$ is an arbitrary reference potential
that is assumed to be independent of $\RMM$.
Using the following Hellman-Feynman theorem for $\EeffQM$,
\BEA
\label{partial_charge}
    \frac
		{ \del \EeffQM( \Rvec, \Phivec' ) }
		{ \del \Phivec' }
	& = &
	\langle \Psi( \Rvec, \Phivec' ) | \Qopvec | \Psi( \Rvec, \Phivec' ) \rangle
    \nonumber
    \\
	& \equiv &
	\Qvec( \Rvec, \Phivec' )
    ,
\EEA
and introducing the internal QM energy as
\BE
\label{EQM}
    \EQM( \Rvec, \Phivec' )
	=
	\langle \Psi( \Rvec, \Phivec' ) | \HQM | \Psi( \Rvec, \Phivec' ) \rangle
    ,
\EE
or alternately via
\BE
\label{internal_QM_energy}
    \EeffQM( \Rvec, \Phivec' )
	=
	\EQM( \Rvec, \Phivec' ) + \Qvec( \Rvec, \Phivec' ) \cdot \Phivec'
    ,
\EE
\Eq{1st_order_expansion} may be rewritten as
\BE
    \EeffQM( \Rvec, \PhivecMM )
	\simeq
	\EQM( \Rvec, \Phivecref )
	+
	\Qvec( \Rvec, \Phivecref )
	\cdot
	\PhivecMM
    .
\EE
(see Appendix \ref{sec:non-var} for cases where the Hellmann-Feynman
theorem does not hold).
Inserting the above expansion of $\EeffQM$
into the exact $\AQMMM( \Rvec )$
and extracting $\EQM( \Rvec, \Phivecref )$
from the integral over $\RMM$, we obtain
\BE
\label{AQMM_ref}
	\AQMMM( \Rvec )
	\simeq
	\EQM( \Rvec, \Phivecref )
	+
	\AMM( \Rvec, \Qvecref )
\EE
with $\Qvecref = \Qvec( \Rvec, \Phivecref )$.
The above equation is very similar to $\AMF( \Rvec )$ in \Eq{AMF_var},
and indeed, the latter can be
recovered simply by setting $\Phivecref$ 
to the self-consistent potential $\PhivecSC$:
\BE
\label{AMF}
    \AMF( \Rvec )
	=
	\EQM( \Rvec, \PhivecSC )
	+
	\AMM( \Rvec, \QvecSC )
    .
\EE
Therefore, the variational principle in \Sec{MF_var} is essentially
equivalent with the first-order expansion
of the effective QM energy about the self-consistent reference field.

%...................................................................

The analytical gradient of $\AMF( \Rvec )$ can be obtained
by first writing $\AMF( \Rvec )$ in terms of the effective QM energy
as
\BEA
    \AMF( \Rvec )
	& = &
	\EeffQM( \Rvec, \PhivecSC( \Rvec ) )
	-
	\QvecSC( \Rvec ) \cdot \PhivecSC( \Rvec )
\nonumber
\\
	& & +
	\AMM( \Rvec, \QvecSC( \Rvec ) )
\EEA
[cf. \Eq{internal_QM_energy}],
taking the $\Rvec$ derivative of the right-hand side,
and using the following symmetric relations:
\BSUB
\BE
    \left.
	  \frac
	  { \del \EeffQM( \Rvec, \Phivec' ) }
	  { \del \Phivec' }
	\right|
	_{ \Phivec' = \PhivecSC }
	=
	\QvecSC( \Rvec )
    ,
\EE
\BE
    \left.
	  \frac
	  { \del \AMM( \Rvec, \Qvec' ) }
	  { \del \Qvec' }
	\right|_{ \Qvec' = \QvecSC }
	=
	\PhivecSC( \Rvec )
    .
\EE
\ESUB
It then follows that the derivatives of $\QvecSC( \Rvec )$ and
$\PhivecSC( \Rvec )$ cancel with each other and we are left with
the following:
\BW
\BE
\label{grad}
    \frac
		{ \del }
		{ \del \Rvec }
	\AMF( \Rvec )
	=
    \left.
	  \frac
	  { \del \EeffQM( \Rvec, \Phivec' ) }
	  { \del \Rvec }
	\right|
	_{ \Phivec' = \PhivecSC }
	+
    \left.
	  \frac
	  { \del \AMM( \Rvec, \Qvec' ) }
	  { \del \Rvec }
	\right|_{ \Qvec' = \QvecSC }
    .
\EE
\EW
This is our working equation for the gradient of QM/MM
free energy with mean-field embedding
[see also \Eqs{grad_cont}{grad_mixed} for related equations].
The first term is the partial $\Rvec$ derivative of the effective
QM energy (\textit{not} of internal QM energy $\EQM$),
which may be written using the Hellman-Feynman theorem as
\BE
\label{EeffQM_deriv}
    \left.
	    \frac
	    { \del \EeffQM( \Rvec, \Phivec' ) }
	    { \del \Rvec_\siteA }
	\right|
	_{ \Phivec' = \PhivecSC }
	=
	\langle \PsiSC |
	    \frac{ \del \HQM }{ \del \Rvec_\siteA }
		+
        \sum_\siteB
		\frac{ \del \Qop_\siteB }{ \del \Rvec_\siteA }
		\espSC_\siteB
	| \PsiSC \rangle
    .
\EE
The second term in \Eq{grad} is the partial $\Rvec$ derivative
of the (classical) solvation free energy, which may be written
using \Eq{AMM} as
\BE
\label{AMM_deriv}
    \left.
    \frac
		{ \del \AMM( \Rvec, \Qvec' ) }
		{ \del \Rvec_\siteA }
    \right|
    _{ \Qvec' = \QvecSC }
    =
	\ll
		Q^\SC_\siteA
		\frac
			{ \del v_{\MM,\siteA} }
			{ \del \Rvec_\siteA }
		+
		\frac
			{ \del \EeffMMvdw }
			{ \del \Rvec_\siteA }
	\gg
    _{ \QvecSC }
    .
\EE

%...................................................................

An alternative way to obtain the free energy
gradient in \Eq{grad} is as follows (see also \Appendix{QMMM_MFEP}).
First we define the mean-field approximation to
the total energy $E( \Rvec, \RMM )$ and
the QM/MM free energy $A( \Rvec )$ as
\BE
\label{EMF}
    \EMF( \Rvec, \RMM )
    =
    \EeffQM( \Rvec, \PhivecSC )
    +
    \QvecSC
    \cdot
    ( \PhivecMM - \PhivecSC )
    +
    \EeffMM
    ,
\EE
and
\BE
    \AMF( \Rvec )
	=
	- \frac{ 1 }{ \beta }
	\ln
	\int d\RMM
	\exp
	\{
	  - \beta
      \EMF( \Rvec, \RMM )
	\}
    .
\EE
The gradient of $\AMF( \Rvec )$ then becomes
\BEA
    \frac
		{ \del }
		{ \del \Rvec }
	\AMF( \Rvec )
    & = &
    \left\langle
      \frac
      { \del \EMF( \Rvec, \RMM ) }
      { \del \Rvec }
    \right\rangle
    _{ \EMF }
\nonumber
\\
    & = &
    \ll
      \frac
      { \del \EMF( \Rvec, \RMM ) }
      { \del \Rvec }
    \gg
    _{ \QvecSC }
    .
\EEA
Inserting \Eq{EMF} into the above equation and using the self-consistency
condition in \Eq{SC_cond} gives the free energy gradient in \Eq{grad}.
We note that if the reference potential $\Phivecref$
is not set at the self-consistent one,
i.e. $\Phivecref \neq \PhivecSC( \Rvec )$,
the following term appears due to incomplete cancellation among terms,
\BE
	\frac
		{ \del \Qvec^\circ( \Rvec ) }
		{ \del \Rvec }
    \cdot
    \left[
	  \ll \PhivecMM \gg_{ \Qvecref }
	  -
	  \Phivecref
	\right]
    ,
\EE
which requires the derivative of QM charges calculated in the
reference potential,
$\del \Qvec^\circ( \Rvec ) / \del \Rvec
= \del \Qvec( \Rvec, \Phivecref ) / \del \Rvec$.

%...................................................................

We now compare the above perturbative approach with
the QM/MM-MFEP\cite{Yang_QMMM_MFEP07,Yang_QMMM_MFEP08}
and ASEP/MD methods.\cite{Aguilar_ASEP_Opt,Aguilar_ASEP_Program}
The QM/MM-MFEP method develops
a series of polarizable QM models
by Taylor expanding its energy and ESP charges
up to first or second order.
Their comparison with the present approach is made
in \Appendix{QMMM_MFEP}. From this comparison it follows that
$\AMF( \Rvec )$ is essentially equivalent with
Model 3 of the QM/MM-MFEP method with charge response kernel $\chi$
neglected.
The free energy gradient of Model 3 (with $\chi$ neglected)
looks somewhat different from the present result at first sight.
However, the former can be rewritten
using the self-consistency condition
as follows (see \Appendix{QMMM_MFEP} for the notation)
\BE
    \rderiv{ A( \rqm ) }
    \simeq
    \rderiv{ \Eeffref }
    +
    \left\langle
      \rderiv{ \EeffMM( \rqm, \rmm ) }
    \right\rangle_{ \tilde{E} }
    ,
\label{Yang_grad}
\EE
which is essentially equivalent
with the present gradient expressions
[Eqs.~(\ref{grad}), (\ref{grad_cont}), and (\ref{grad_mixed})].

%...................................................................

The above equation provides some rationale
for the approximate gradient used by the ASEP/MD method
[\Eq{Aguilar_grad}]. To see this, let us rewrite \Eq{Yang_grad} as
\BE
\label{Aguilar_grad_detail}
    \rderiv{ A( \rqm ) }
    \simeq
    \rderivTot
    \langle
    \Psiref
    |
      \HQM
      +
      \VintMF
    |
    \Psiref
    \rangle
    ,
\EE
with
\begin{multline}
    \VintMF
    =
    \MMave
    [
      \int d\xvec
      \rhoop( \xvec )
      \vmm( \xvec, \rmmref( \tau ) )
\\
      +
      \EeffMM( \rqm, \rmmref( \tau ) )
    ]
    ,
\end{multline}
which suggests that the gradient of $A( \rqm )$
may be viewed as the gradient of effective QM energy
calculated in the averaged MM potential.
Here it should be noted that the $\rqm$-derivative above does not act
on $\Psiref$ nor $\rmmref( \tau )$, no matter whether
the self-consistency condition is assumed or not
(see \Appendix{mixed_rep} for details).

%-----------------------------------------------------------------------------

\subsection{\label{sec:fluc}Statistical fluctuations of the QM wavefunction}

As seen above, the variational/mean-field approach
totally neglects statistical fluctuations of the QM wavefunction
about the mean-field state. The aim of this section is thus to
discuss several ways for evaluating such non-mean-field effects
on QM/MM free energy.

%...................................................................

First, let us separate the total energy
into the mean-field and non-mean-field contributions as follows:
\BE
\label{E_separate}
    E( \Rvec, \RMM )
    =
    \EMF( \Rvec, \RMM )
    +
    \Efluc( \Rvec, \RMM )
    ,
\EE
where $\EMF( \Rvec, \RMM )$ is defined by \Eq{EMF}
and $\Efluc( \Rvec, \RMM )$ is the remaining part of the total energy.
Using the definition of
$E( \Rvec, \RMM )$ in \Eq{total_energy},
the non-mean-field term can be written more explicitly as
\begin{multline}
    \Efluc( \Rvec, \RMM )
    =
    \EeffQM( \Rvec, \PhivecMM )
\\
    -
    \EeffQM( \Rvec, \PhivecSC )
    -
    \QvecSC
    \cdot
    ( \PhivecMM - \PhivecSC )
    .
\end{multline}
Inserting \Eq{E_separate} into the exact $A( \Rvec )$
in \Eq{AQMMM} gives
\BE
\label{A_separate}
    \AQMMM( \Rvec )
    =
    \AMF( \Rvec )
    +
    \Afluc( \Rvec )
    ,
\EE
where
\BE
\label{Afluc}
    \Afluc( \Rvec )
	=
	- \frac{ 1 }{ \beta }
	\ln
	\ll
	\exp( - \beta \Efluc )
	\gg
	_{ \QvecSC }
    .
\EE
We note that up to this point \Eqs{A_separate}{Afluc} are still exact.
The statistical average in \Eq{Afluc}
can be evaluated rather rigorously as follows.
First, one calculates a long trajectory
of the MM subsystem
using the sampling function $\exp( - \beta \EMF )$,
selects a relatively small subset of MM configurations
from the long trajectory  (say, 500 samples),
and calculates $\Efluc$ for those selected
configurations in order to take the average of $\exp( -\beta \Efluc )$.
Indeed, this is a type of dual-level QM/MM sampling method,
where $\exp( - \beta \EMF )$ is used as a low-cost sampling function
while $\Efluc$ gives energetic corrections.

%...................................................................

Although the above dual-level method is rigorous,
it requires hundreds of QM calculations and thus
may be rather expensive. One approach for reducing the computational cost
is to truncate the expansion of effective QM energy
at the second order,\cite{Naka_RISM_Fluc,Yang_QMMM_RPP,Yang_QMMM_MFEP08}
\BE
\label{Efluc_approx}
    \Efluc
    \simeq
    \Efluc^{(2)}
	=
	\half
	( \PhivecMM - \PhivecSC )
	\cdot \chiQM \cdot
	( \PhivecMM - \PhivecSC )
    ,
\EE
where $\chiQM$ is defined by
\BE
\label{chiQM}
    \chiQM( \Rvec, \Phivec' )
	=
	\frac
		{ \del^2 \EeffQM( \Rvec, \Phivec' ) }
		{ \del \Phivec' \del \Phivec' }
	=
	\frac
		{ \del \Qvec( \Rvec, \Phivec' ) }
		{ \del \Phivec' }
\EE
with $\Phivec' = \PhivecSC$. 
$\chiQM$ is also called the charge response kernel
due to the second equality in \Eq{chiQM}.\cite{Morita_CRK1,Morita_CRK2}
Once $\chiQM$ is obtained,
the statistical average of $\exp( - \beta \Efluc^{(2)} )$ can be
evaluated with no extra QM calculations, thus significantly reducing
the computational cost.
The above second-order expansion is also
utilized by Models 2 and 3 of the QM/MM-MFEP method
[see \Eq{Yang_E_model3} in the present paper] in order to
describe statistical fluctuations of the QM wavefunction.
\cite{Yang_QMMM_MFEP07,Yang_QMMM_MFEP08}

%...................................................................

A further simplification can be made by introducing a
Gaussian fluctuation model for the MM environment.
Specifically, we assume that the
MM electrostatic potential acting on QM atoms,
$\PhivecMM = \PhivecMM( \Rvec, \RMM )$, takes
a (multi-dimensional) Gaussian distribution:
\cite{Levy_GaussFluc,Levy_Review_GaussFluc,Simonson_GaussFluc}
\begin{multline}
\label{GF}
    \ll
	\delta( \Phivec' - \PhivecMM )
	\gg
	_{ \QvecSC }
\\
    \propto
	\exp
	\left[
	  - \half
	  ( \Phivec' - \PhivecSC )
	  \cdot \sigmaMM^{-1} \cdot
	  ( \Phivec' - \PhivecSC )
	\right]
    .
\end{multline}
Here, $\sigmaMM$ is the covariance matrix of $\PhivecMM$,
\BE
    \sigmaMM
	=
	\ll
	( \PhivecMM - \PhivecSC )
	( \PhivecMM - \PhivecSC )^T
	\gg
	_{ \QvecSC }
    .
\EE
By combining $\Efluc^{(2)}$ in \Eq{Efluc_approx}
and the Gaussian fluctuation model above,
we obtain an approximate analytical expression for $\Afluc( \Rvec )$:
\BE
\label{Afluc_GF}
    \Afluc( \Rvec )
	\simeq
	\frac{ 1 }{ 2\beta }
	\ln
	\det
	[
	  1 + \beta \chiQM \sigmaMM
	]
    .
\EE
Note that since $\chiQM$ is negative definite,
\cite{Morita_CRK1}
$\Efluc^{(2)}$ and $\Afluc( \Rvec )$ are always negative.
The basic appeal of \Eq{Afluc_GF} is that
once the charge response kernel $\chiQM$ is obtained,
$\Afluc( \Rvec )$ can also be obtained simultaneously
by combining with $\sigmaMM$ that is available
from the mean-field calculation.
$\chiQM$ can be evaluated most efficiently
by solving a coupled-perturbed
Hartree-Fock or Kohn-Sham equation,
\cite{Morita_CRK1,Morita_CRK2,Morita_CRK_DFT}
or more primitively, by numerically differentiating
the ESP charges $\Qvec( \Rvec, \Phivec' )$ with respect
to $\Phivec'$ based on the second equality in \Eq{chiQM}.

%=============================================================================

%- \clearpage

\section{\label{sec:app}Application to an \SNtwo reaction in water}

%-----------------------------------------------------------------------------

\subsection{\label{sec:background}Background}

We now apply the above method to a Type-II \SNtwo reaction in water
(the Menshutkin reaction)
\BE
\label{Menshutkin}
    \mathrm{
      NH_3 + CH_3Cl
      \rightarrow
      NH_3CH_3^{+} + Cl^{-}
    }
    .
\EE
This reaction is known to exhibit greatly enhanced rates
in polar solvents than in the gas phase
due to strong electrostatic stabilization of
the products.\cite{Menshutkin,Reinchardt_Book}
This is in contrast to Type-I \SNtwo reactions like
$
    \mathrm{
      Cl^{-} + CH_3Cl
      \rightarrow
      ClCH_3 + Cl^{-}
    }
    ,
$
which are decelerated by greater
electrostatic stabilization of the reactant than
of the transition state.
Due to the great acceleration in rate,
the Menshutkin reaction became the subject
of many theoretical studies.\cite{Sola_Mensh,Gao_Mensh,GaoXia_Mensh,Shaik_Mensh,Rivail_Mensh,Truong_Mensh,Mennucci_Mensh,Wiberg_Mensh,Gordon_Mensh,Naka_Mensh,Nagaoka_Mensh,Sola_Mensh_Pair,Aguilar_ASEP_Mensh,Moliner_QMMM_IC_Mensh,Shaik_Mensh_VB,Higashi_LRFE}
Gao and Xia performed the first extensive
QM/MM study using the AM1 model,\cite{GaoXia_Mensh}
and demonstrated that the transition state in water
is shifted remarkably toward the reactant region.
Continuum solvent models were also applied to the same reaction
at various levels of QM methods.
\cite{Rivail_Mensh,Truong_Mensh,Mennucci_Mensh}
While those studies observed that continuum models can provide free energetics
similar to the QM/MM results,\cite{GaoXia_Mensh}
it was also argued that those models
may not be appropriate for reaction (\ref{Menshutkin})
due to the presence of hydrogen bonds.\cite{Wiberg_Mensh}
Since then, several QM/MM(-type) calculations were performed,
\cite{Wiberg_Mensh,Gordon_Mensh,Naka_Mensh,Nagaoka_Mensh,Sola_Mensh_Pair,Aguilar_ASEP_Mensh,Higashi_LRFE,Moliner_QMMM_IC_Mensh} including
the RISM-SCF method,\cite{Naka_Mensh}
a mean-field QM/MM approach,\cite{Aguilar_ASEP_Mensh}
and a dual-level method.\cite{Moliner_QMMM_IC_Mensh}
Overall, those calculations are in reasonable agreement
with each other, predicting the free energy of activation
$\GTS$ to be 20 $\sim$ 30 kcal/mol and
the free energy of reaction $\Greact$ to be $-20$ $\sim$ $-35$
kcal/mol (both including solute entropic contributions).
Among those studies, the present one is most similar in spirit
to the mean-field QM/MM calculation by Aguilar and co-workers.
\cite{Aguilar_ASEP_Mensh}

%-----------------------------------------------------------------------------

\subsection{\label{sec:details}Computational details}

Following previous studies, we define the reaction coordinate as
\BE
\label{RC}
    \RC( \Rvec ) = r( \mathrm{C-Cl} ) - r( \mathrm{C-N} )
    .
\EE
The mean-field free energy $\AMF( \Rvec )$ is
minimized with respect to $\Rvec$ under the
constraint $\RC( \Rvec ) = s'$. The resulting optimized geometry
will be denoted as $\Ropt( \RC' )$.
Our goal here is to obtain the free energy profile
$\AMF( \Ropt( \RC' ) )$ as a function
of $\RC'$. In this paper we constructed such a profile
by integrating $\nabla \AMF$ along the optimized reaction path
$\Ropt( \RC' )$ [i.e., via thermodynamic integration (TI)]:
\begin{multline}
\label{TI}
    \AMF( \Ropt( \RC_b ) ) - \AMF( \Ropt( \RC_a ) )
\\
	=
	\int^{ \RC_b }_{ \RC_a } d \RC'
	\frac{ \del \Ropt( \RC' ) }{ \del \RC' }
	\cdot
	\nabla \AMF( \Ropt( \RC' ) )
    ,
\end{multline}
where $\nabla = \del / \del \Rvec$. We calculated
$\Ropt( s' )$ and $\nabla \AMF( \Ropt( s' ) )$ for equally spaced grid points,
$\RC_{k} = 0.2 k$ \AA~$(k = 0, \pm 1, \ldots)$, and evaluated
the above integral via cubic spline interpolation.
In practice, the following trapezoid rule
was also sufficient for this small size of grid spacing:
\begin{multline}
\label{TI_trapezoid}
    \AMF( \Ropt( \RC_{K} ) ) - \AMF( \Ropt( \RC_{0} ) )
\\
	\simeq
	\sum_{ k = 1, K }
	\left[
	  \Ropt( \RC_{k} ) - \Ropt( \RC_{k-1} )
	\right]
\qquad\qquad\qquad
\\
\qquad
	\times
	\frac{1}{2}
	\left[
	\nabla \AMF( \Ropt( \RC_{k} ) )
	+
	\nabla \AMF( \Ropt( \RC_{k-1} ) )
	\right]
    .
\end{multline}

%...................................................................

The geometry optimization on $\AMF( \Rvec )$ was performed
by adapting the sequential sampling/optimization method by
Yang and co-workers\citeYang for the present purpose.
Since the free energy gradient in \Eq{grad}
assumes that the self-consistency (SC) condition in \Eq{SC_cond}
is satisfied, one might think that the latter must be solved for each step
of the optimization. Then, the optimization procedure may appear
the following:
\begin{enumerate}
\item
Given a geometry $\Rvec^{(n)}$ at an intermediate step $n$,
solve the SC condition in \Eq{SC_cond}
for obtaining $\PhivecSC$ and $\QvecSC$,
and evaluate the analytical gradient
$\nabla \AMF( \Rvec^{(n)} )$ via \Eq{grad};
\item
Advance $\Rvec^{(n)}$ one step, e.g., as
$\Rvec^{(n+1)} := \Rvec^{(n)} - \lambda \nabla \AMF( \Rvec^{(n)} )$; and
\item
Repeat steps 1 and 2 until a given convergence criterion is met,
e.g., $| \nabla \AMF( \Rvec^{(n)} ) | < \epsilon$.
\end{enumerate}
However, this scheme is somewhat too restrictive because
the gradient does not have to be exact nor
very accurate at the early stages of the optimization.
What is needed is that the gradient becomes increasingly more accurate
as the optimization proceeds.
This observation leads to the following
variant of the sequential sampling/optimization
procedure,\citeYang which
performs the statistical sampling of the MM environment and
the optimization of the QM geometry in an iterative manner:
\begin{enumerate}
\item Cycle 0: \textit{QM optimization.} \\
The QM subsystem is optimized in the gas phase to prepare the initial state.
The resulting QM geometry and partial charges are denoted
as $\Rvec^{(0)}$ and $\Qvec^{(0)}$.
No MM/MD simulation is performed at this cycle.
\item Cycle $n$: (i) \textit{MM sampling.} \\
The QM geometry $\Rvec^{(n-1)}$ and charges $\Qvec^{(n-1)}$ of
the previous cycle are embedded into the MM environment.
An MM/MD simulation is then performed to evaluate
the averaged electrostatic potential
and the gradient of $\AMM$ in \Eq{AMM_deriv}:
\BE
    \Phivecave^{(n)}
    =
    \ll \PhivecMM \gg
    _{ \Rvec^{(n-1)}, \Qvec^{(n-1)} }
    ,
\EE
\BEA
    \AMMgrad^{(n)}_\siteA
    & = &
	\left.
	  \frac
	  { \del \AMM( \Rvec, \Qvec' ) }
	  { \del \Rvec_\siteA }
	\right|_{ \Rvec^{(n-1)}, \Qvec^{(n-1)} }
    \\
    & = &
	\ll
      Q^{(n-1)}_\siteA
      \frac
      { \del v_{\MM,\siteA} }
      { \del \Rvec_\siteA }
      +
      \frac
      { \del \EeffMMvdw }
      { \del \Rvec_\siteA }
	\gg
    _{ \Rvec^{(n-1)}, \Qvec^{(n-1)} }
    .
    \nonumber
\EEA
Since $\Phivecave^{(n)}$ and $\AMMgrad^{(n)} = 
(\AMMgrad^{(n)}_1, \ldots, \AMMgrad^{(n)}_N)$
are simple
$N$- and $3N$-dimensional vectors ($N$ is the number of QM atoms),
they can be accumulated directly in the MD simulation.
\item Cycle $n$: (ii) \textit{QM optimization.} \\
The QM geometry is optimized in the presence of
$\Phivecave^{(n)}$ and $\AMMgrad^{(n)}$.
To this end, we employ the following
target function for optimizing the QM geometry $\Rvec$:
\BEA
\label{target_E}
    A^{(n)}( \Rvec )
    & = &
	\EeffQM( \Rvec, \Phivecave^{(n)} )
    \nonumber
    \\
	& & +
    \sum_\siteA
	\AMMgrad^{(n)}_\siteA \cdot ( \Rvec_\siteA - \Rvec^{(n-1)}_\siteA )
    .
\EEA
The gradient of this target function is
\BEA
\label{grad_approx}
    \frac{ \del }{ \del \Rvec_\siteA }
    A^{(n)}( \Rvec )
	& = &
    \frac
		{ \del \EeffQM( \Rvec, \Phivecave^{(n)} ) }
		{ \del \Rvec_\siteA }
	+
	\AMMgrad^{(n)}_\siteA
\\
	& = &
	\left.
	  \frac
	  { \del \EeffQM( \Rvec, \Phivec' ) }
	  { \del \Rvec_\siteA }
	\right|_{ \Phivec' = \Phivecave^{(n)} }
\nonumber
\\
	& & +
	\left.
	  \frac
	  { \del \AMM( \Rvec, \Qvec' ) }
	  { \del \Rvec_\siteA }
	\right|_{ \Rvec^{(n-1)}, \Qvec^{(n-1)} }
    .
\nonumber
\EEA
We note that $A^{(n)}( \Rvec )$ 
gives a local approximation to
$\AMF( \Rvec )$ in that its gradient
$\nabla A^{(n)}( \Rvec )$ approximates
the analytical gradient $\nabla \AMF( \Rvec )$ in \Eq{grad}.
By minimizing $A^{(n)}( \Rvec )$
under the constraint $s( \Rvec ) = s'$ [i.e., by eliminating
the orthogonal component of $\nabla A^{(n)}( \Rvec )$
to $\nabla s( \Rvec )$],
we obtain a new QM geometry $\Rvec^{(n)}$ for the next cycle.
In this paper the optimization of $A^{(n)}( \Rvec )$ was
performed by adding linear external potentials
$\AMMgrad_\siteA \cdot ( \Rvec_\siteA - \Rvec_\siteA^{(n-1)} )$
and forces $-\AMMgrad_\siteA$ for individual QM atoms
in the GAMESS quantum chemistry package.\cite{Gamess}
\item By iterating over the above cycles, 
the QM geometry, ESP charges, and MM mean potentials
converge to their asymptotic values,
namely $\Rvec^{(n)} \approx \Rvec^{(n-1)}$, $\Qvec^{(n)} \approx \Qvec^{(n-1)}$,
and $\Phivec^{(n)} \approx \Phivec^{(n-1)}$.
This means that the SC condition is satisfied to
a good accuracy. Since the QM geometry does not move
any further, we may regard $\nabla A^{(n)}( \Rvec^{(n)} )$
as providing a good approximation to $\nabla \AMF( \Ropt )$
at the optimized geometry $\Ropt \simeq \Rvec^{(n)}$.
\end{enumerate}
The above iterative optimization was used
to obtain the free energy gradient in \Eq{TI}.

%...................................................................

The other computational details are as follows.
The QM and MM calculations were performed using
modified versions of GAMESS\cite{Gamess}
and DL\_POLY packages.\cite{Dlpoly}
Following Truong \etal{},\cite{Truong_Mensh}
we used the BHHLYP/6-31+G(d,p) method in most calculations.
Previous study shows that this method gives results similar to
the MP4/aug-cc-pVDZ level for the present reaction.\cite{Truong_Mensh}
The ESP charge operator and associated
fitting grid were defined following
Ten-no \etal{}\cite{Tenno_RISM}~and Spackman.\cite{Spackman,Note_ESP_grid}
The charge response kernel $\chiQM$
was calculated via finite difference
of the ESP charges $\Qvec( \Rvec, \Phivec' )$
with respect to $\Phivec'$.

%...................................................................

$C_{3v}$ symmetry was enforced on the QM subsystem. The optimization tolerance
was set to $5\times10^{-4}$ hartree/bohr, which is 
five times larger than the default setting in GAMESS.
Although $\Phivecave^{(n)}$ and $\AMMgrad^{(n)}$ above
should satisfy $C_{3v}$ symmetry in principle,
they do not in practice due to statistical errors.
These errors generate small artificial components of overall
translation and rotation. We thus removed
those components manually such that the optimization
could be completed to a given tolerance.

%...................................................................

MD calculations were performed by solvating one solute molecule
into 253 water molecules (with the TIP3P potential)\cite{TIP3P}
in a cubic box of side length 19.7 \AA{}
at $T$ = 300 K.
Periodic boundary condition was applied, and electrostatic
potentials were calculated using the Ewald method.
The Lennard-Jones parameters are listed in Table \ref{tab:LJ}.
The timestep for integration was 2 fs.
One iterative optimization cycle consisted of
50000 MD steps for equilibration and
300000 steps for production.
Although 100000 production steps were sufficient
for obtaining a similar result,
we did not attempt to minimize the computational efforts.
Rather, we aimed at obtaining a highly converged result,
such that the statistical errors become
comparable to the width of the plotting line.

%...................................................................

\begin{table}
\caption{\label{tab:LJ}Lennard-Jones parameters for the solute molecule.
Parameters of the Cl atom are taken from Gao and Xia (\Ref{GaoXia_Mensh}),
and those of the other atoms are from the AMBER94 force field (\Ref{Amber94}).
}
\begin{ruledtabular}
\begin{tabular}{ccc}
Atom  &  $\sigma$ (\AA)  & $\epsilon$ (kcal/mol)  \\
\hline
C               &  3.3996  &  0.1094  \\
N               &  3.3409  &  0.1700  \\
H$_\mathrm{C}$  &  2.4713  &  0.0157  \\
H$_\mathrm{N}$  &  1.0691  &  0.0157  \\
Cl              &  4.1964  &  0.1119  \\
\end{tabular}
\end{ruledtabular}

\end{table}

%-----------------------------------------------------------------------------

\subsection{\label{sec:profiles}Free energy profiles}

To illustrate the above optimization procedure,
\Fig{grad_cycle} displays the $z$ component of the approximate free energy
gradient $\nabla A^{(n)}( \Rvec^{(n)} )$ in \Eq{grad_approx}
as a function of iterative optimization cycle $n$.
(Here the solute molecule was kept oriented in the $z$ direction of the
simulation box, so only the $z$ component of the gradient is nonvanishing.)
As seen, the approximate gradients converge monotonically
to their asymptotic values as cycle $n$ proceeds.
Other quantities like $\Qvec^{(n)}$ and $\Phivec^{(n)}$
exhibited a similar convergence behavior. We thus expect
that the self-consistency is achieved to a good
accuracy in the last few cycles of the iterative procedure.
In this paper we used 8 cycles for each value of the reaction
coordinate $s$. \Figure{grad} plots the gradients thus obtained
as a function of $s$.

%...................................................................

\BFIG
    % Fig.1
    \caption{ \label{fig:grad_cycle}
      The $z$ component of approximate free energy gradient
      $\nabla A^{(n)}$ (in kcal/mol/\AA) at $s = 0.0$ \AA~
      as a function of iterative optimization cycle $n$.
    }
    \includegraphics[width=7.5cm,clip]{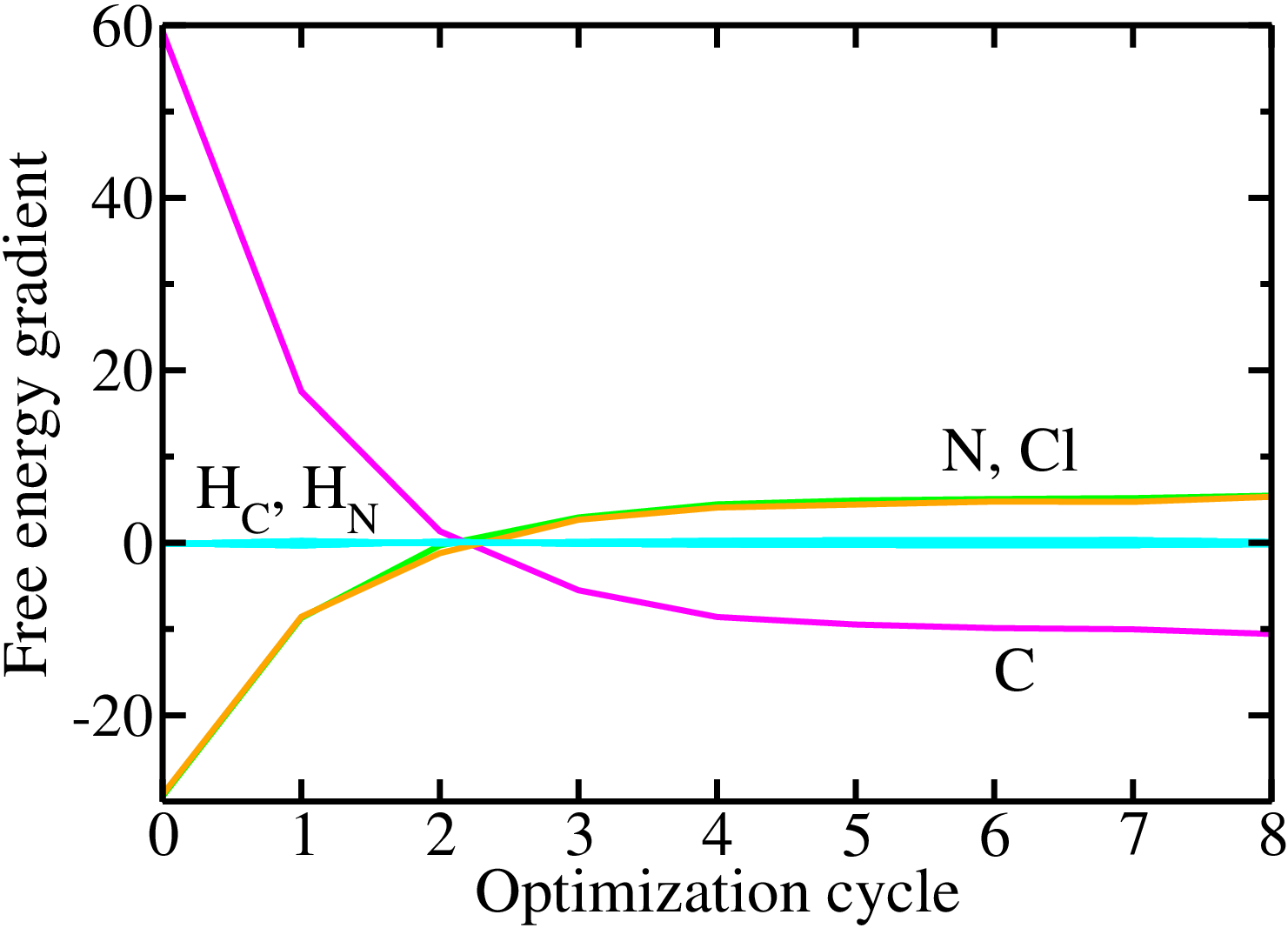}
\EFIG

\BFIG
    % Fig.2
    \caption{ \label{fig:grad}
      The $z$ components of free energy gradient $\nabla \AMF( \Ropt( s ) )$
      (in kcal/mol/\AA) as a function of reaction coordinate $s$ (in \AA).
      The arrow indicates the location of the transition state in solution.
    }
    \includegraphics[width=7.5cm,clip]{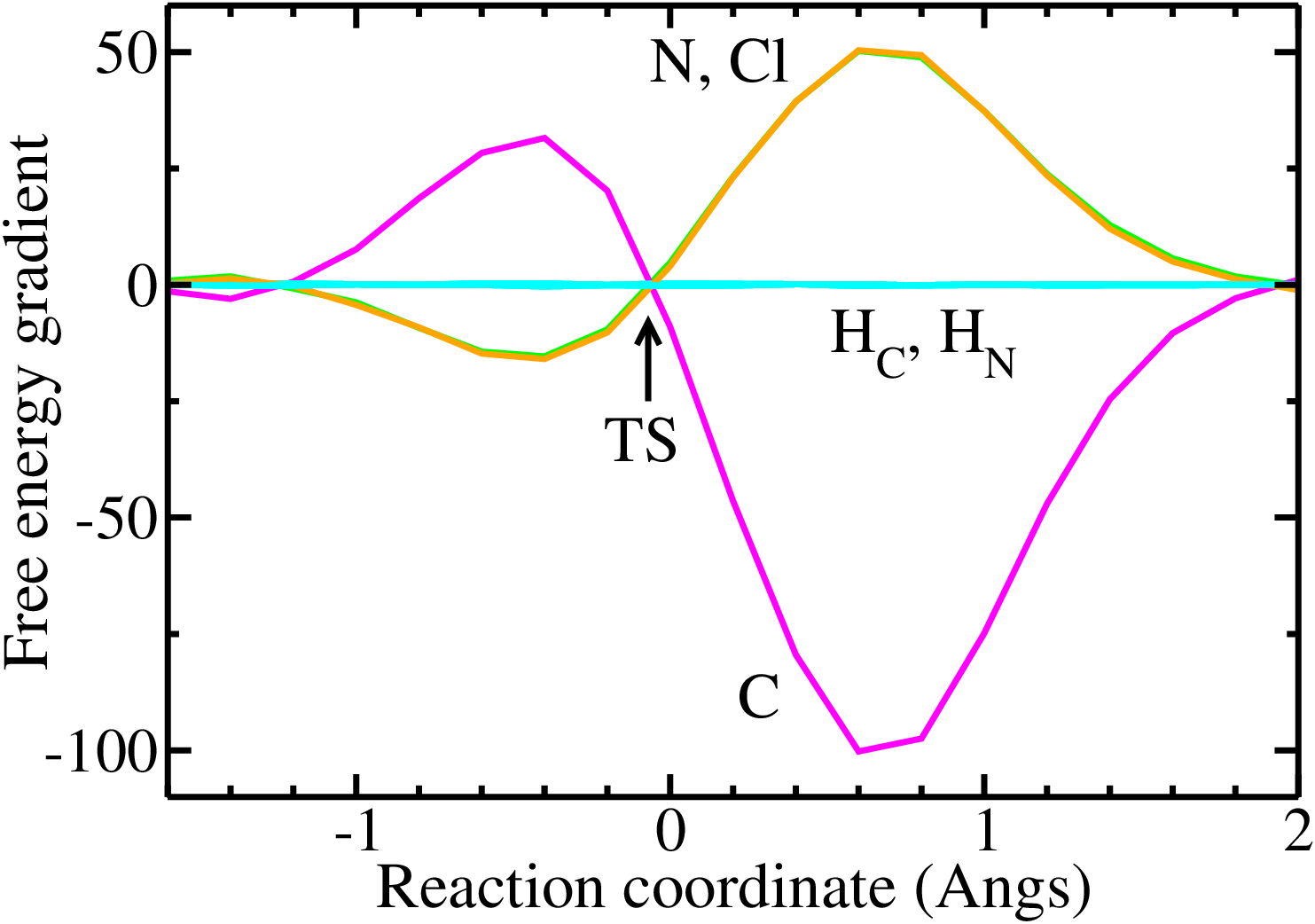}
\EFIG

%...................................................................

By integrating the gradients in \Fig{grad},
we obtain a free energy profile $\AMF(s) \equiv \AMF(\Rvecopt(s))$
in \Fig{profile} (solid line with circles).
The barrier top of $\AMF(s)$ is located at $s^\ddagger = -0.05$ \AA,
which corresponds to $\rCNTS = 2.215$ \AA~ and $\rCClTS = 2.165$ \AA.
The free energy of activation and of reaction are defined here as
\BEA
    \ATS & = & \AMF( s^\ddagger ) - \AMF( s= -1.6 )
    ,
    \nonumber
    \\
    \Areact & = & \AMF( s = 2.0 ) - \AMF( s = -1.6 )
    ,
    \nonumber
\EEA
which are found to be
$\ATS = 10.6$ kcal/mol and $\Areact = -38.7$ kcal/mol
at the BHHLYP/6-31+G(d,p) level.
By adding solute entropic contributions,\cite{Truong_Mensh,Mennucci_Mensh}
we obtain $\GTS = \ATS + 13.1 = 23.7$ kcal/mol,
which is in good agreement with $\GTS = 25.6$ kcal/mol
obtained by Aguilar \etal{}~at the BHHLYP/aug-cc-pVDZ level.
\cite{Aguilar_ASEP_Mensh}
On the other hand, the reaction free energy is
$\Greact = \Areact + 7.5 = -31.2$ kcal/mol,
which falls within the error bar of the experimental result,
$-34\pm10$ kcal/mol.\cite{GaoXia_Mensh}

%...................................................................

\BFIG
    % Fig.3
    \caption{ \label{fig:profile}
      Free energy profile $\AMF$ in \Eq{AMF} at the BHHLYP/6-31+G(d,p) level
      without solute entropic contributions (solid line with circles).
      The solid line and circles are obtained with \Eqs{TI}{TI_trapezoid},
      respectively.
      $\EQM$, $\EQM(\mathrm{gas})$, and $\AMM$ represent
      the internal QM energy in \Eq{EQM}, its gas-phase counterpart,
      and the (relative) solvation free energy in \Eq{AMM}, respectively.
      All the profiles are depicted such that
      they coincide at $s = -1.6$ \AA.
    }
    \includegraphics[width=7.5cm,clip]{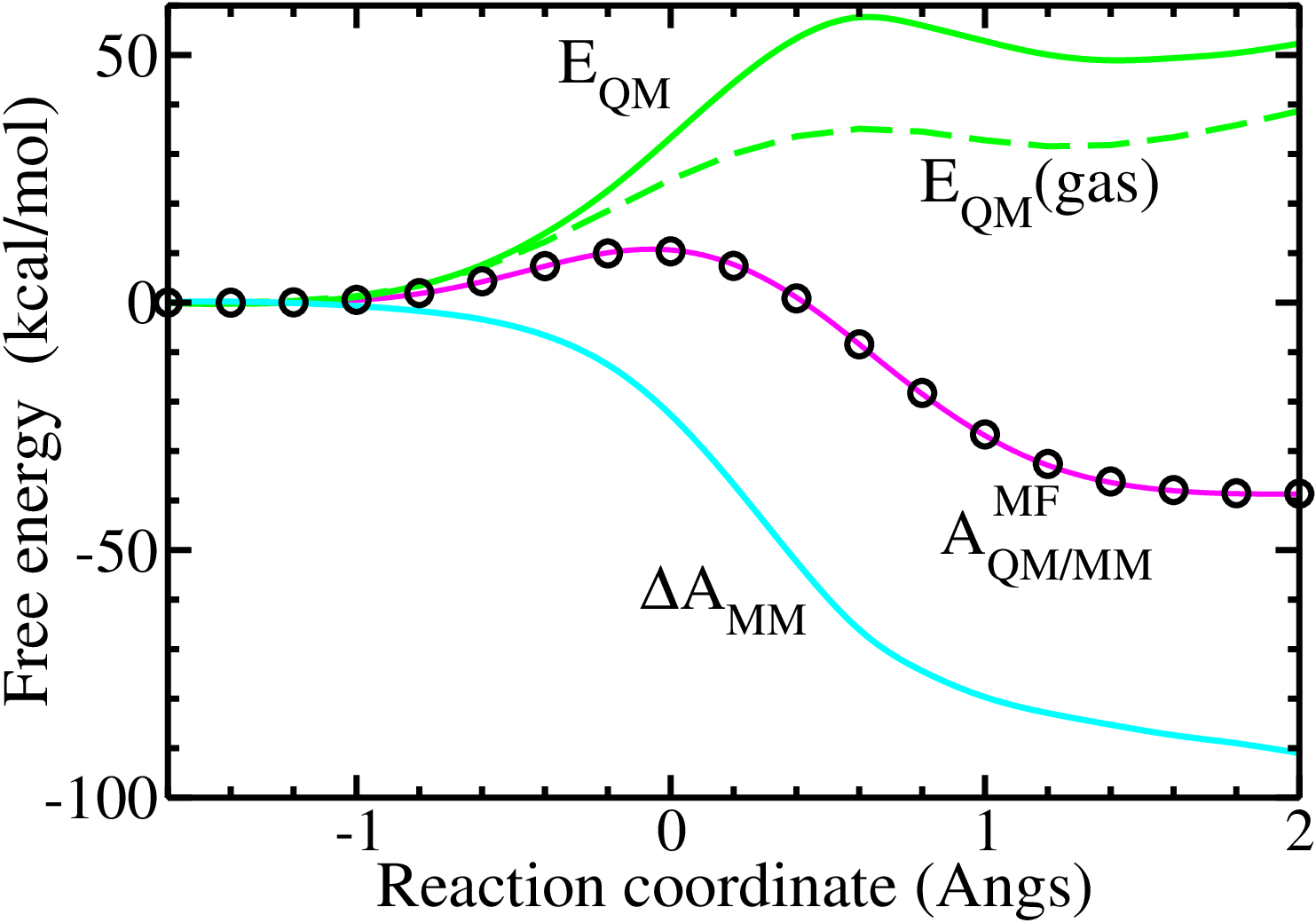}
\EFIG

%...................................................................

\Figure{profile} also illustrates how the internal QM energy
$\EQM( \Rvec, \PhivecSC )$ and the (relative) solvation free energy
$\AMM( \Rvec, \QvecSC )$ vary as functions of $s$.
The gas-phase counterpart of the former [i.e.,
$\EQM( \Rvec, \Phivec' = \mathbf{0} )$ along the gas-phase optimized
path] is also plotted.
To facilitate the comparison, all the profiles are shifted vertically
such that they coincide at $s = -1.6$ \AA.
\Figure{profile} shows that $\AMF$ is determined by
strong cancellation between $\EQM$ and $\AMM$.
While the QM electronic energy increases steeply
with the separation of the ion pair,
this is more than compensated by strong electrostatic stabilization
by the solvent. Figures \ref{fig:charge} and \ref{fig:vmm} illustrate
how the QM fragment charges and
MM mean potentials vary as functions of $s$.
The optimized reaction paths in the gas phase and in solution
are compared in \Fig{react_path}.
As stressed previously,\cite{GaoXia_Mensh}
the transition state in solution [i.e., the saddle point of $\AMF( \Rvec )$]
is shifted remarkably toward the reactant region.
This indicates that for the present charge separation reaction,
the transition state in the gas phase should not be used
for calculating the activation free energy in solution.

%...................................................................

\BFIG
    % Fig.4
    \caption{ \label{fig:charge}
      Fragment charges for $\mathrm{CH_3}$, $\mathrm{NH_3}$,
      and Cl atom in solution (solid lines)
      and in the gas phase (dashed lines).
    }
    \includegraphics[width=7.5cm,clip]{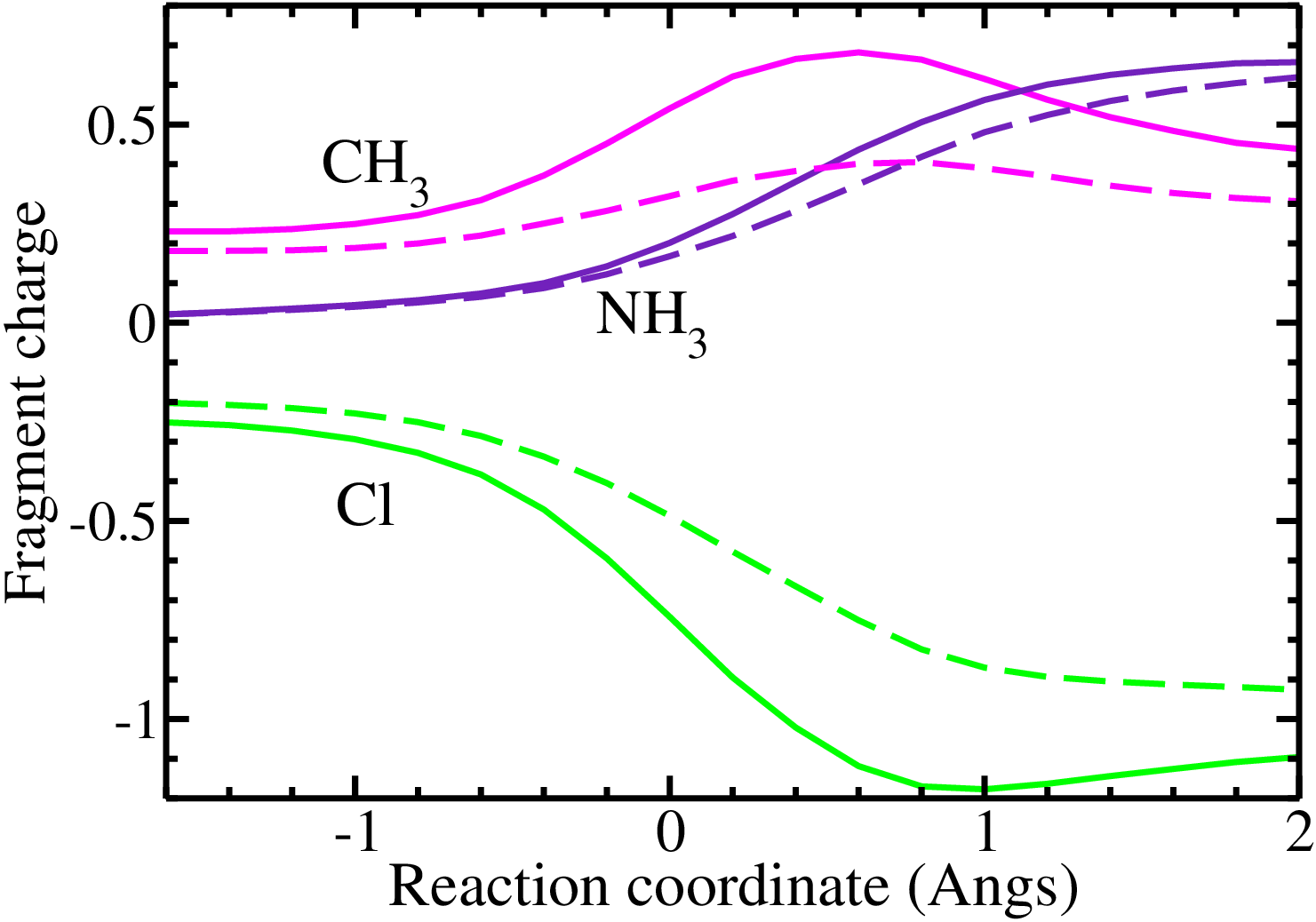}
\EFIG

\BFIG
    % Fig.5
    \caption{ \label{fig:vmm}
      Mean electrostatic potentials $\PhivecSC$ from the MM environment.
    }
    \includegraphics[width=7.5cm,clip]{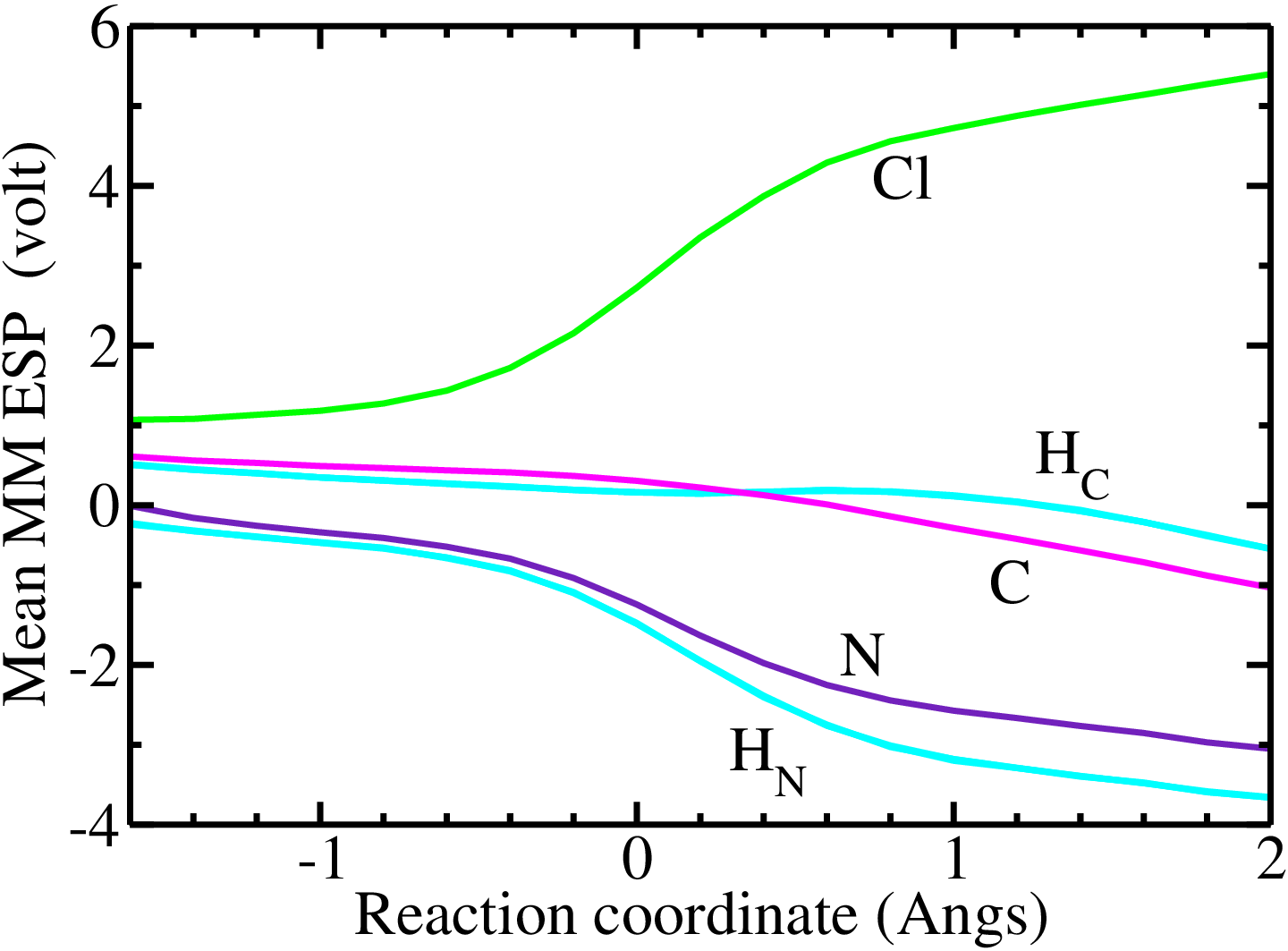}
\EFIG

\BFIG
    % Fig.6
    \caption{ \label{fig:react_path}
      Reaction paths optimized in solution (solid line)
      and in the gas phase (dashed line).
      ``TS(aq)'' and ``TS(gas)'' represent the transition states
      corresponding to the top of the barrier of
      $\AMF$ and $\EQM(\mathrm{gas})$ in \Fig{profile}, respectively.
    }
    \includegraphics[width=7.5cm,clip]{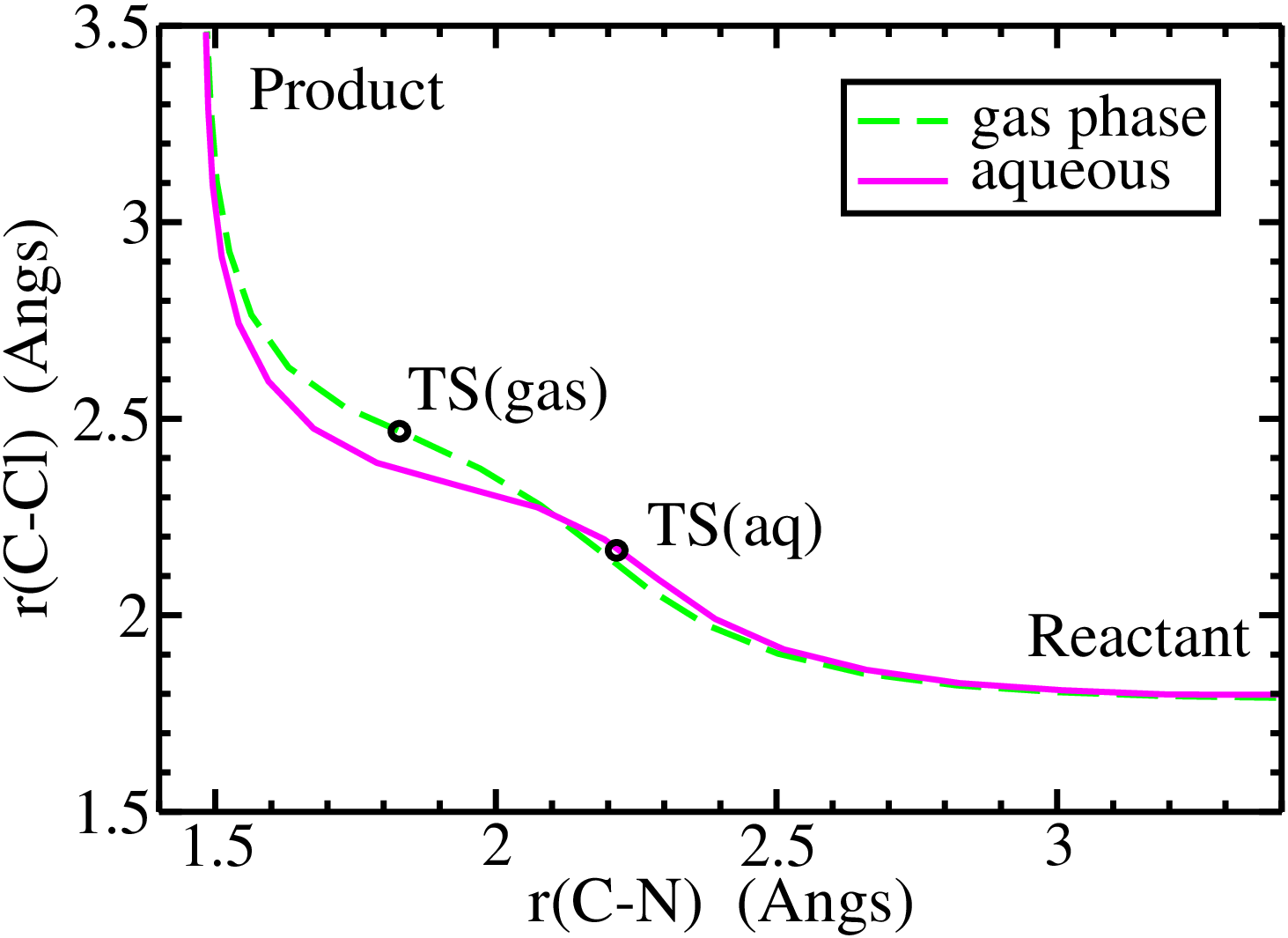}
\EFIG

%...................................................................

To check the validity of the free energy gradient,
separate free-energy perturbation (FEP) calculations were also
performed. Free energy differences between neighboring points of
$s_k$ (corresponding to circles in \Fig{profile})
were calculated as
\begin{multline}
    \AMF( \Ropt_{ k+1 } )
    -
    \AMF( \Ropt_{ k } )
\\
    =
    -
    \frac{ 1 }{ \beta }
    \ln
    \ll
    \exp(
      - \beta
      \Delta E_{ k+1,k }
    )
    \gg
    _{ k }
    ,
\end{multline}
where
$\Ropt_{ k } = \Ropt( s_{ k } )$,
$
  \Delta E_{ k+1,k }
  =
  \EMF( \Ropt_{ k+1 }, \RMM )
  -
  \EMF( \Ropt_{ k   }, \RMM )
$ with $\EMF( \Rvec, \RMM )$ given in \Eq{EMF},
and $\ll \cdots \gg_{ k }$ denotes the statistical average
with the sampling function $\exp[ - \beta \EMF( \Ropt_{ k }, \RMM ) ]$.
The necessary input like $\Ropt_{ k }$ was obtained from the TI calculation.
Since FEP does not utilize the gradient information,
the comparison of FEP profiles with TI ones
offers a stringent test of consistency between
$\AMF( \Rvec )$ and $\nabla \AMF( \Rvec )$.
\Figure{fep} shows that the FEP profiles thus obtained
are in excellent agreement with the TI ones, indicating
that the free energy gradient is calculated correctly.
Although there are slight differences between the FEP
and TI profiles in the product region, this may be due to
electrostatic finite-size effects,\cite{Note_FEP_error}
because the agreement becomes better for a larger number of
solvent molecules $N = 997$ than $N = 253$.

%...................................................................

\BFIG
    % Fig.7
    \caption{\label{fig:fep}
      Free energy profiles obtained with thermodynamic
      integration (TI) and free
      energy perturbation (FEP).
      $N$ is the number of water solvent molecules.
    }
    \includegraphics[width=7.5cm,clip]{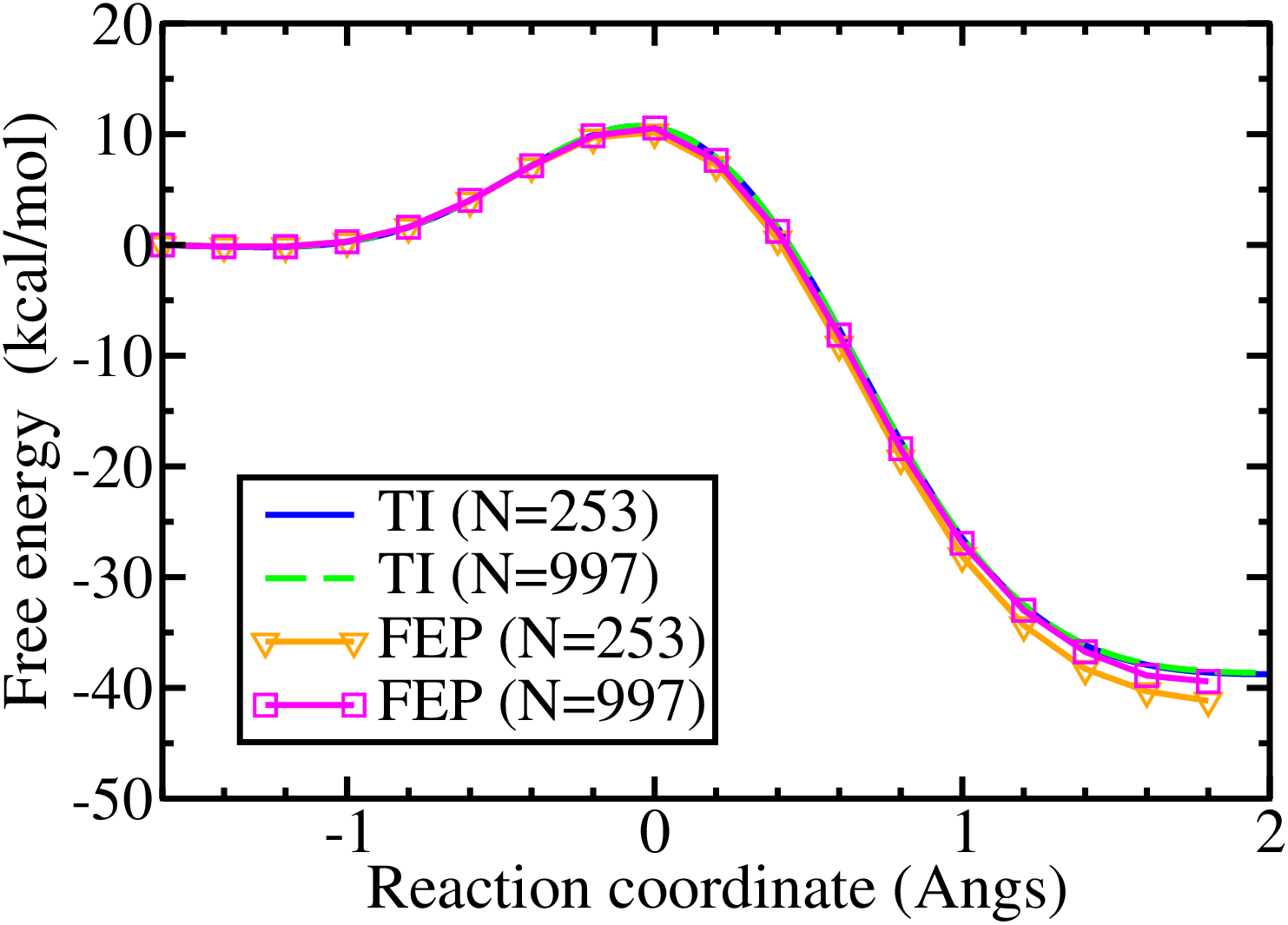}
\EFIG

%...................................................................

\Figure{profile_compare} compares the free energy profiles obtained at the
HF, MP2, B3LYP, and BHHLYP levels with a larger basis set 6-311+G(2d,2p).
This figure shows that the MP2 gives the highest value
of the free energy barrier, $\ATS = 15.5$ kcal/mol,
the B3LYP gives the lowest value, 9.2 kcal/mol, and
the BHHLYP their intermediate, 11.9 kcal/mol (without including
solute entropic contributions). \Table{free_energy} summarizes those
values of $\ATS$ and $\Areact$ obtained with various
QM methods and basis sets. If we assume that the BHHLYP
gives the ``best'' energetics
for the present reaction,\cite{Truong_Mensh}
our main results are
$\GTS = 11.9 + 13.1 = 25.0$ kcal/mol
and $\Greact = -37.1 + 7.5 = -29.6$ kcal/mol (including
solute entropic contributions).

%...................................................................

\BFIG
    % Fig.8
    \caption{\label{fig:profile_compare}
      Free energy profiles obtained at the HF,
      MP2, B3LYP, and BHHLYP levels with the 6-311+G(2d,2p) basis set.
      The MP2 gives the highest value
      of $\ATS$, while the B3LYP gives the lowest.
      See \Table{free_energy} for individual values
      of $\ATS$ and $\Areact$.
      The profiles do not include solute entropic contributions.
      The RESP method is used for all calculations.
    }
    \includegraphics[width=7.5cm,clip]{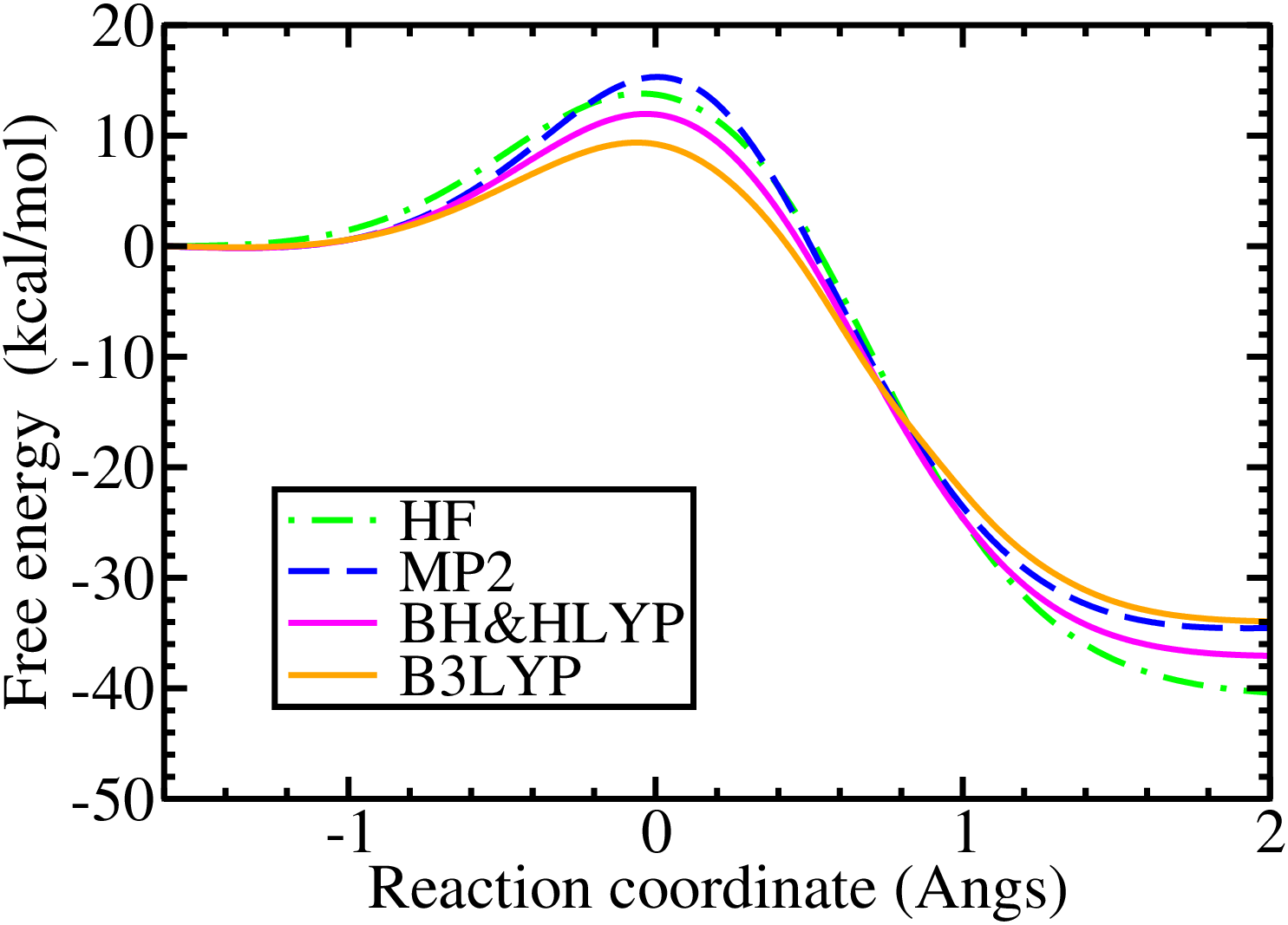}
\EFIG

%...................................................................

\begin{table}
\caption{\label{tab:free_energy}
Free energy of activation $\ATS$ and
reaction $\Areact$ (in kcal/mol) obtained at the HF, MP2, B3LYP,
and BHHLYP levels with the 6-31+G(d,p) basis.
Values in parentheses are obtained with the 6-311+G(2d,2p) basis.
To compare with the previous studies, one needs to add
solute entropic contributions to $\ATS$ and $\Areact$ such that
$\GTS \simeq \ATS + 13.1$ and $\Greact \simeq \Areact + 7.5$ kcal/mol
(\Refs{Truong_Mensh}{Mennucci_Mensh}).
The RESP method is used for all calculations unless otherwise noted.
}
\begin{ruledtabular}
\begin{tabular}{ccc}
Method  &  $\ATS$  &  $\Areact$  \\
\hline
HF     &   12.0  (13.7)   &   $-41.6$  ($-40.4$)  \\
MP2    &   16.8  (15.5)   &   $-35.1$  ($-34.6$)  \\
B3LYP  &    7.8  (9.2)   &   $-36.9$   ($-33.9$)  \\
BHHLYP &   10.6  (11.9)   &   $-38.7$  ($-37.1$)  \\
BHHLYP\footnotemark[1]
       &   10.6  (11.9)   &   $-38.7$  ($-34.0$)  \\
\end{tabular}
\end{ruledtabular}
\footnotetext[1]{RESP method not used.}
\end{table}

%...................................................................

To estimate the non-mean-field effects on QM/MM free energy,
\Table{Afluc} lists the values of $\Afluc$
evaluated using the Gaussian fluctuation model in \Eq{Afluc_GF}.
This table also gives the values of $\EeffQMfluc$ defined by
\BE
\label{Efluc_def}
    \EeffQMfluc
    =
    \ll \EeffQM( \Rvec, \PhivecMM ) \gg_{ \QvecSC }
    -
    \EeffQM( \Rvec, \PhivecSC )
    ,
\EE
which was also calculated using the Gaussian model as
\BE
\label{Efluc_GF}
    \EeffQMfluc
    \simeq
    \frac{ 1 }{ 2 }
    \mathrm{tr}[ \chiQM \sigmaMM ]
    .
\EE
This quantity was used previously by Naka \etal{}\cite{Naka_RISM_Fluc}
and Aguilar \etal{}\cite{Aguilar_ASEP_Stark} in order to study
fluctuations of the QM wavefunction in solution.
The table shows that the absolute values of $\Afluc$ and $\EeffQMfluc$
are considerably small ($<0.5$ kcal/mol)
for the entire region of the reaction coordinate. They are also similar
to the values reported
for other organic molecules in water ($\EeffQMfluc = -0.2 \sim -0.5$ kcal/mol).
\cite{Naka_Mensh,Aguilar_ASEP_Stark}
It should be noted that the impact of $\Afluc( \Rvec )$ on
free energy profiles is even smaller,
because the variation of $\Afluc( \Ropt( s ) )$ as a function of $s$
is on the order of 0.1 kcal/mol.
This result suggests that the non-mean-field effects on QM/MM free energy
are rather small for the present reaction in water.
Similar observations have been made in the literature.
\cite{Aguilar_ASEP_Stark,Aguilar_ASEP_Compare,Thiel_QMMM_FEPvsTI,Takahashi_QMMM_FrozenDensity}
Nevertheless, we stress that
it is not clear at present to what extent
this conclusion applies to different types of systems, e.g.,
enzyme reactions where
local fluctuations of the MM environment may deviate
significantly from the Gaussian distribution.\cite{Simonson_GaussFluc}

%...................................................................

\begin{table}
\caption{\label{tab:Afluc} Fluctuation corrections for the free energy
$\Afluc$ in \Eq{Afluc_GF} and for the interaction energy
$\EeffQMfluc$ in \Eq{Efluc_GF} calculated with the BHHLYP/6-31+G(d,p) method.
Values in parentheses are obtained with the 6-311+G(2d,2p) basis.
RC stands for the reaction coordinate in \Eq{RC}.
All energies are given in kcal/mol.
}
\begin{ruledtabular}
\begin{tabular}{cccc}
RC \ (\AA)  &  $\Afluc$  &  $\EeffQMfluc$  \\
\hline
  $-1.6$  &  $-0.43$ ($-0.44$)  &  $-0.38$ ($-0.39$)  \\
  $ 0.0$  &  $-0.41$ ($-0.45$)  &  $-0.36$ ($-0.40$)  \\
  $ 2.0$  &  $-0.30$ ($-0.34$)  &  $-0.28$ ($-0.31$)  \\
\end{tabular}
\end{ruledtabular}
\end{table}

%...................................................................

\section{\label{sec:concl}Discussions and conclusions}

\textit{Numerical stability of the ESP charge operator.}
ESP charges and associated charge operator $\Qvecop$ are
sometimes numerically unstable, as often stressed
in the literature.\cite{Yang_Charge,Yokogawa_Charge,Morita_CRK_DFT}
For example, we observed
an oscillatory behavior of partial
charges within the CH$_3$ group during the optimization cycles.
This was typical for $s = 0.8$ \AA, where the ion pair products
start to form. Since these oscillations are partly due to
ambiguous assignment of partial charges for ``buried'' atoms,
\cite{Morita_CRK_DFT}
the RESP method\cite{RESP} was of great help in suppressing those
oscillations. However, the RESP method was of little help
in removing a divergent behavior of partial
charges within the NH$_3$ group observed in the reactant asymptotic region
($s < -2.0$ \AA). Specifically, the partial charge on the N (H$_\mathrm{N}$) atom
kept on growing in the negative (positive) direction during the optimization
cycles.
This might be due to inherent limitations of the present charge model,
where partial charges are placed only on atomic nuclei and
the lone pair on the N atom may be poorly described.
\cite{Aguilar_ASEP_Compare}
In this respect, it may be more straightforward
to use the continuous or mixed representation
in \Appendix{cont_rep} or \ref{sec:mixed_rep},
where one embeds the MM point charges directly into the QM Hamiltonian.
See \Refs{Aguilar_ASEP_Program}{Yang_QMMM_MFEP08}
for this type of implementation.

%...................................................................

\textit{FEP that connects optimized geometries.}
If one is interested only in the free energy difference between
two stationary points (e.g., activation free energy),
it is probably more efficient to use FEP than TI.
Specifically, one first searches the free energy surface
for stationary points by using the free energy gradient
(and possibly the hessian), and then connects these points
via FEP. The QM geometries and charges of intermediate points could be
generated by linear interpolation of two end points.
See \Ref{Aguilar_ASEP_Mensh} for such a calculation.
In this way, one can reduce the number of costly free energy optimization.
If one also needs to know a rough free energy
profile, one could perform additional optimization
for a limited number of intermediate points and then connect
them via FEP.

%...................................................................

\textit{Solute thermal/entropic contributions.}
The method in this paper calculates the QM/MM free energy for a given
fixed QM geometry. The thermal/entropic contributions
of the QM subsystem thus need to be taken into account separately,
e.g., via harmonic vibration approximations.
This is a well-known limitation of the present type of method,
which is also shared by conventional solvation theories.
To overcome this limitation, several methods have been proposed
for \apriori{} including the solute flexibility
into the QM/MM free energy calculation at a reasonable computational cost.
\cite{Warshel_QMMM_DualLevel06,Yang_QMMM_RPP,Higashi_MCMM}

%...................................................................

To conclude, we have presented a direct QM/MM analog of conventional
solvation theories based on variational and perturbative frameworks.
The main approximation in this paper is
that the true QM wavefunction is replaced by an averaged one
that is calculated in the MM mean field.
We stress however that the electrostatic interactions
between the averaged QM wavefunction and the MM environment are calculated
correctly without further approximations.
The basic appeal of the mean-field QM/MM approach
is that it can describe different environments (e.g., solutions and enzymes)
on an equal theoretical footing,
while the number of QM calculations can be made significantly smaller than
a direct QM/MM calculation.

%=============================================================================

\begin{acknowledgments}

This work was supported by Grant-in-Aid for the Global COE Program,
"International Center for Integrated Research and Advanced Education
in Materials Science," from the Ministry of Education, Culture, Sports,
Science and Technology of Japan. The author also thanks Prof.~Weitao Yang
for a critical reading of the manuscript and suggesting detailed comparison
with the QM/MM-MFEP method.

\end{acknowledgments}

%=============================================================================

\appendix

\section{\label{sec:cont_rep}Continuous representation}

The main text is based on the approximate \Schroedinger equation in \Eq{SE},
where QM/MM electrostatic interactions are ``discretized'' in terms of
the ESP charge operator.
In this section we summarize an alternative
formulation using the continuous \Schroedinger equation in \Eq{SE_cont}.

%...................................................................

First, the total energy is given by
\BE
    E( \Rvec, \RMM )
    =
	\EeffQM[ \Rvec, \espMM ]
	+
	\EeffMMvdw( \Rvec, \RMM )
    ,
\EE
where $\EeffQM[ \Rvec, \espMM ]$ is defined via \Eq{SE_cont} with
$v'( \xvec ) = \vmm( \xvec, \RMM )$.
We then expand $\EeffQM[ \Rvec, \espMM ]$
in terms of $\espMM(\xvec,\RMM )$ up to first order,
\BW
\BEA
    \EeffQM[ \Rvec, \espMM ]
	& \simeq &
	\EeffQM[ \Rvec, \espSC ]
	+
	\int d\xvec
	\rhoSC( \xvec | \Rvec )
    \{
	  \espMM( \xvec, \RMM  )
      -
      \espSC( \xvec | \Rvec )
    \}
    \nonumber
    \\
    & = &
	\EQM[ \Rvec, \espSC ]
	+
	\int d\xvec
	\rhoSC( \xvec | \Rvec )
	\espMM( \xvec , \RMM  )
    ,
\EEA
\EW
where
$\EQM[ \Rvec, \espSC ] = \langle \PsiSC | \HQM | \PsiSC \rangle$
and we have used the following Hellmann-Feynman theorem:
\BE
    \frac
    { \delta \EeffQM[ \Rvec, v' ] }
    { \delta v'( \xvec ) }
    =
    \langle \Psi[ \Rvec, v' ] |
    \rhoop( \xvec )
    | \Psi[ \Rvec, v' ] \rangle
    .
\EE
$\rhoSC( \xvec | \Rvec )$ and
$\espSC( \xvec | \Rvec )$ are obtained from the following self-consistency
condition,
\BSUB
\label{SC_cond_cont}
\BE
\label{vSC_cont}
    \espSC( \xvec | \Rvec )
	=
	\ll \espMM( \xvec , \RMM ) \gg_{ \rhoSC }
    ,
\EE
\BE
	\rhoSC( \xvec | \Rvec )
	=
	\langle \PsiSC | \rhoop( \xvec ) | \PsiSC \rangle
    ,
\EE
\ESUB
with $\PsiSC \equiv \Psi[ \Rvec, \espSC ]$.
Note that $\espSC$ and $\rhoSC$ depend parametrically on $\Rvec$
via \Eq{SC_cond_cont}.
$\ll \cdots \gg$ is the statistical average defined by
\BE
    \ll \cdots \gg_{ \rho' }
    =
    \frac
    { \int d\RMM e^{ - \beta \Eeff[ \rho', \Rvec, \RMM ] } (\cdots) }
    { \int d\RMM e^{ - \beta \Eeff[ \rho', \Rvec, \RMM ] } }
    ,
\EE
with
\BE
\label{EeffMM_cont}
    \Eeff[ \rho', \Rvec, \RMM ]
	=
	\int d \xvec
	\rho'( \xvec )
	\espMM( \xvec, \RMM )
	+
	\EeffMMvdw( \Rvec, \RMM )
    .
\EE
Inserting the above first-order expansion
into $\AQMMM( \Rvec )$ in \Eq{AQMMM} gives the QM/MM free energy
with mean-field embedding,
\BE
\label{AMF_cont}
    \AMF( \Rvec )
	=
	\EQM[ \Rvec, \espSC ]
	+
	\AMM[ \Rvec, \rhoSC ]
    ,
\EE
with
\BE
\label{AMM_cont}
    \AMM[ \Rvec, \rho' ]
	=
	\intMM
	e^{
	  - \beta \Eeff[ \rho', \Rvec, \RMM ]
	}
    .
\EE
%
%...................................................................
%
The gradient of $\AMF( \Rvec )$ can be obtained via similar arguments as
\BE
\label{grad_cont}
    \frac
		{ \del }
		{ \del \Rvec }
	\AMF( \Rvec )
	=
    \left.
	  \frac
	  { \del \EeffQM[ \Rvec, \esp' ] }
	  { \del \Rvec }
	\right|
	_{ \esp' = \espSC }
	+
    \left.
	  \frac
	  { \del \AMM[ \Rvec, \rho' ] }
	  { \del \Rvec }
	\right|_{ \rho' = \rhoSC }
    .
\EE
The first term represents the energy
gradient in a fixed external field.
The second term may be rewritten using \Eq{AMM_cont} as
\BE
\label{AMM_deriv_cont}
    \left.
	  \frac
	  { \del \AMM[ \Rvec, \rho' ] }
	  { \del \Rvec }
	\right|_{ \rho' = \rhoSC }
    = \;
	\ll
	  \frac
	  { \del \EeffMMvdw( \Rvec, \RMM ) }
	  { \del \Rvec }
	\gg
	_{ \rhoSC }
    .
\EE
Note that the above equation lacks the electrostatic term
like $ Q^\SC_\siteA \ll \del v_{\MM,\siteA} / \del \Rvec_\siteA \gg$
that is present in the the discretized case [\Eq{AMM_deriv}].
This discrepancy originates from the different physical meaning of
$\del \EeffQM[ \Rvec, \esp' ] / \del \Rvec $
(in the continuous representation)
and $\del \EeffQM( \Rvec, \Phivec' ) / \del \Rvec$
(in the discretized representation).
In the discretized case,
the external potential values $\Phivec'$ acting on QM atoms
are kept constant while varying the nuclear coordinates $\Rvec$.
In the continuous case, the external potential \textit{field}
$\esp'( \xvec )$ is kept constant
while varying $\Rvec$. This means that
the potential values acting on QM atoms,
$\esp'( \Rvec_\siteA )$, may vary as a function of $\Rvec$.
The situation becomes clear by considering the following relation:
\BEA
    \left.
	  \frac
	  { \del \EeffQM[ \Rvec, \esp' ] }
	  { \del \Rvec_\siteA }
	\right|
	_{ \esp' = \espSC }
    & \simeq &
    \left.
	  \frac
	  { \del }
	  { \del \Rvec_\siteA }
	  \EeffQM( \Rvec, \{ \espSC( \Rvec_\siteB | \Xvec ) \} )
    \right|
    _{ \Xvec = \Rvec }
    \nonumber
	\\
	& \simeq &
    \left.
	  \frac
	  { \del \EeffQM( \Rvec, \Phivec' ) }
	  { \del \Rvec_\siteA }
	\right|
	_{ \Phivec' = \PhivecSC }
    \\
	& + &
	Q^\SC_\siteA( \Rvec )
	\ll
	  \frac
	  { \del \espMM( \Rvec_\siteA, \RMM ) }
	  { \del \Rvec_\siteA }
	\gg
	_{ \QvecSC }
    ,
    \nonumber
\EEA
where we have used the following identity obtained from \Eq{vSC_cont}:
\BEA
\label{vmm_deriv}
    \left.
    \frac
	{ \del \espSC( \Rvec_\siteA | \Xvec ) }
	{ \del \Rvec_\siteA }
    \right|
    _{ \Xvec = \Rvec }
    & = &
    \left.
    \frac
	{ \del \espSC( \xvec | \Rvec ) }
	{ \del \xvec }
    \right|
    _{ \xvec = \Rvec_\siteA }
\\
    & = &
    \left.
	\ll
	  \frac
	  { \del \espMM( \xvec, \RMM ) }
	  { \del \xvec }
	\gg
    _{ \rhoSC }
    \right|_{ \xvec = \Rvec_\siteA }
    .
\nonumber
\EEA
Therefore, it follows that
the missing electrostatic term in \Eq{AMM_deriv_cont}
is now accounted for by the energy gradient term in \Eq{grad_cont}.

%-----------------------------------------------------------------------------

\section{\label{sec:ESP_charge}ESP charge operator}

The ESP charges $\{ Q_\siteA \}$ for a given wavefunction $\Psi$ are
obtained by minimizing the following function\cite{RESP}
\BEA
\label{ESP_L}
    L( \Qvec, \lambda )
    & = &
    \sum_l^\mathrm{grid}
    \left\{
      \int d\xvec
      \frac
      { \langle \Psi | \rhoop( \xvec ) | \Psi \rangle }
      { | \Gvec_l - \xvec | }
      -
      \sum_\siteA
      \frac
      { Q_\siteA }
      { | \Gvec_l - \Rvec_\siteA | }
    \right\}^2
\nonumber
\\
    & & -
    \lambda
    \left[
      \sum_\siteA
      Q_\siteA
      -
      \Qtot
    \right]
    ,
\EEA
where $\{ \Gvec_l \}$ are the ESP fitting grid and $\Qtot$ is
the total charge.
By requiring that $\del L( \Qvec, \lambda ) / \del \Qvec = 0$,
inverting the resulting linear equations for $\Qvec$,
and determining $\lambda$ via $\sum_\siteA Q_\siteA = \Qtot$, we obtain
\BE
    Q_\siteA
    =
    \langle \Psi |
      \Qop_\siteA
    | \Psi \rangle
    ,
\EE
where $\Qop_\siteA$ is an explicit function of
$
        \{ \rvec_i \},
        \{ \Rvec_\siteA \},
        \{ \Gvec_k \},
$
and $\Qtot$,
with $\{ \rvec_i \}$ being the electron coordinates.
See previous work for the explicit form of $\Qop_\siteA$
in the atomic orbital basis.
\cite{Tenno_RISM,Sato_RISM,Morita_CRK1,Morita_CRK2,Morita_CRK_DFT}
The above definition of $\Qvecop$ suggests that one may make
the following replacement
\BE
  \int d\xvec
  \frac
  { \rhoop( \xvec ) }
  { | \yvec - \xvec | }
  \simeq
  \sum_\siteA
  \frac
  { \Qop_\siteA }
  { | \yvec - \Rvec_\siteA | }
  ,
\EE
as long as $\yvec$ is located
outside the core region of the QM charge density.
Then, the continuous QM/MM electrostatic interaction may be discretized as
\BE
    \int d\xvec
    \rhoop( \xvec )
    \vmm( \xvec, \RMM )
    \simeq
    \sum_\siteA
    \Qop_\siteA
    \vmm( \Rvec_\siteA, \RMM )
    ,
\EE
by inserting the definition of $\vmm( \xvec, \RMM )$ in \Eq{vmm}.
This is the present rationale for using $\Qvecop$ in \Eq{SE}.

%-----------------------------------------------------------------------------

\section{\label{sec:QMMM_MFEP}Comparison with the QM/MM-MFEP method}

Here we compare the perturbative treatment of $\AMF( \Rvec )$
in \Sec{MF_pert} with the QM/MM-MFEP method.
\cite{Yang_QMMM_MFEP07,Yang_QMMM_MFEP08}
The starting point is the same as \Eq{AQMMM}
(here expressed using the notation in \RefYang),
\BE
\label{Yang_AQMMM}
    \AQMMM( \rqm )
    =
    - \frac{ 1 }{ \beta }
    \ln
    \int
    d \rmm
    \exp\{ - \beta E( \rqm, \rmm ) \}
    ,
\EE
where $\rqm = \Rvec$, $\rmm = \RMM$, and
the total energy is given by
\BE
\label{Yang_E}
    E(\rqm,\rmm)
    =
    \langle \Psi | \Heff | \Psi \rangle
    +
    \EeffMM( \rqm, \rmm )
    .
\EE
$\Heff$ is the effective QM Hamiltonian in the presence of
the MM electrostatic field,
\BE
\label{Yang_Heff}
    \Heff
    =
    \Hqm
    +
    \int d\xvec \rhoop( \xvec ) \vmm( \xvec, \rmm )
    ,
\EE
and other quantities like $\HQM$ are defined in the main text.
We then introduce the MM mean field as
\BE
\label{Yang_vmmref}
    \vmmref( \xvec )
    =
    \MMave
    \vmm( \xvec, \rmmref( \tau ) )
    ,
\EE
where $\{ \rmmref(\tau) \}_{ \tau = 1 }^L$
are a set of MM configurations
obtained from the previous cycle of the sequential sampling/{}optimization
method.\cite{Yang_QMMM_MFEP08}
The QM Hamiltonian in the presence of the MM mean field is defined by
\BE
\label{Yang_Heffref}
    \Heffref
    =
    \Hqm
    +
    \int d\xvec \rhoop( \xvec ) \vmmref( \xvec )
    .
\EE
The eigenfunction and eigenenergy of $\Heffref$ are denoted as
$| \Psi^\circ \rangle$ and
$\langle \Psi^\circ | \Heff^\circ | \Psi^\circ \rangle \equiv \Eeffref$,
and the ESP charges derived from
$| \Psi^\circ \rangle$ are written as $\Qref_i( \rqm )$.
The internal QM energy associated with $\Heffref$ is defined as
\BE
\label{Yang_Einternal}
    \Einternal( \rqm )
    \equiv
    \Eeffref
    -
    \sumqmi
    \Qref_i( \rqm ) \vmmref( \rqmi )
    ,
\EE
i.e., by subtracting the QM/MM electrostatic interaction energy
expressed in terms of ESP charges from the effective QM energy.

%...................................................................

The QM/MM-MFEP method then develops a series of
polarizable QM models by Taylor expanding its energy
and ESP charges up to first or second order.
Among others, Model 3 (``QM point charges
with polarization due to MM and QM atoms'')
approximates the total energy as follows
[Eqs.~(36) and (40) of \RefYang]:
\BEA
    \tilde{E}( \rqm, \rmm )
    & = &
    \Einternal( \rqm )
    \\
    & + &
    \sumqmi
    \Qref_i( \rqm ) \vmm( \rqmi, \rmm )
    \nonumber
    \\
    & + &
	\half
    \sumqmi
    \sumqmj
	[ \vmm( \rqmi, \rmm ) - \vmmref( \rqmi ) ]
    \nonumber
    \\
    & & \times
	\chi_{ij}
	[ \vmm( \rqmj, \rmm ) - \vmmref( \rqmj ) ]
    \nonumber
    \\
    & + &
    \EeffMM( \rqm, \rmm )
    ,
    \nonumber
\label{Yang_E_model3}
\EEA
where $\chi_{ij}$ is the charge response kernel in \Eq{chiQM}.
The above equation may be viewed as the second-order
expansion of the effective QM energy in terms of MM electrostatic
potential [cf. \Eq{Efluc_approx}].
The gradient of QM/MM free energy is
obtained by inserting \Eq{Yang_E_model3} into the following,
\BE
\label{Yang_grad0}
    \rderiv{ A( \rqm )}
    \simeq
    \left\langle
      \rderiv{ \tilde{E}( \rqm, \rmm ) }
    \right\rangle
    _{ \tilde{ E } }
    ,
\EE
or alternately, into an FEP-type expression
[Eq.~(6) of \RefYang]
\BE
\label{Yang_FEP}
    \rderiv{ A( \rqm ) }
    \simeq
    \frac
    {
    \left\langle
      \rderiv{ \tilde{E}( \rqm, \rmm ) }
      e^{ - \beta ( \tilde{E}( \rqm, \rmm ) - \Eref( \rmm ) ) }
    \right\rangle
    _{ \Eref }
    }
    {
    \left\langle
      e^{ - \beta ( \tilde{E}( \rqm, \rmm ) - \Eref( \rmm ) ) }
    \right\rangle
    _{ \Eref }
    }
    ,
\EE
where $\Eref( \rmm )$ is the reference sampling function that is obtained from
the previous cycle of the sequential sampling/{}optimization
method.\citeYang

%...................................................................

The main difference of the present approach
from the QM/MM-MFEP method is that
the present one utilizes the self-consistency
condition in order to simplify the gradient expression.
To see this, let us insert $\tilde{E}(\rqm,\rmm)$ in \Eq{Yang_E_model3}
into the statistical average in \Eq{Yang_grad0},
\BEA
    \rderiv{ A( \rqm ) }
    & = &
    \rderiv{ \Einternal( \rqm ) }
    \nonumber
    \\
    & & +
    \sumqmi
    \rderiv{ \Qref_i( \rqm ) }
    \langle \vmm( \rqmi, \rmm ) \rangle_{ \tilde{E} }
    \nonumber
    \\
    & & +
    \sumqmi
    \Qref_i( \rqm )
    \left\langle
      \rderiv{ \vmm( \rqmi, \rmm ) }
    \right\rangle_{ \tilde{E} }
    \nonumber
    \\
    & & +
    \left\langle
      \rderiv{ \EeffMM( \rqm, \rmm ) }
    \right\rangle_{ \tilde{E} }
    ,
\label{Yang_grad1}
\EEA
where terms depending on $\chi_{ij}$ have been neglected
(they are treated separately in \Sec{fluc}).
Using \Eq{Yang_Einternal}, we may rewrite
the above equation as
\BW
\BEA
    \rderiv{ A( \rqm ) }
    & = &
    \rderiv{ \Eeffref }
    +
    \sumqmi
    \rderiv{ \Qref_i( \rqm ) }
    \left\{
      \langle \vmm( \rqmi, \rmm ) \rangle_{ \tilde{E} }
      -
      \vmmref( \rqmi )
    \right\}
    \nonumber
    \\
    & & +
    \sumqmi
    \Qref_i( \rqm )
    \left\{
      \left\langle
        \rderiv{ \vmm( \rqmi, \rmm ) }
      \right\rangle_{ \tilde{E} }
      -
      \rderiv{ \vmmref( \rqmi ) }
    \right\}
    +
    \left\langle
      \rderiv{ \EeffMM( \rqm, \rmm ) }
    \right\rangle_{ \tilde{E} }
    .
\label{Yang_grad2}
\EEA
\EW
Now let us assume that the reference MM coordinates
$\{ \rmmref( \tau ) \}$ satisfies the following
self-consistency condition
\BE
\label{Yang_SC_cond}
    \langle f( \rqm, \rmm ) \rangle_{ \tilde{E} }
    \simeq
    \MMave
    f( \rqm, \rmmref( \tau ) ),
    \;\;\;  \forall f
    ,
\EE
which is expected to hold well for the last few cycles
of the sequential sampling/{}optimization method.\citeYang
Then, by setting $ f = \vmm( \rqmi, \rmm )$
or $f = \del \vmm( \rqmi, \rmm ) / \del \rqm$, we have
\BSUB
\label{Yang_vmm_deriv}
\BEA
  \langle \vmm( \rqmi, \rmm ) \rangle_{ \tilde{E} }
  & \simeq &
  \MMave
  \vmm( \rqmi, \rmmref( \tau ) )
  \nonumber
  \\
  & = &
  \vmmref( \rqmi )
  ,
  \\
  \left\langle
    \rderiv{ \vmm( \rqmi, \rmm ) }
  \right\rangle_{ \tilde{E} }
  & \simeq &
  \MMave
  \rderiv{ \vmm( \rqmi, \rmmref( \tau ) ) }
  \nonumber
  \\
  & = &
  \rderiv{ \vmmref( \rqmi ) }
  ,
\EEA
\ESUB
which suggest that the curly brackets in \Eq{Yang_grad2}
vanish, and as a result we obtain a simpler expression for the
free energy gradient,
\BE
    \rderiv{ A( \rqm ) }
    \simeq
    \rderiv{ \Eeffref }
    +
    \left\langle
      \rderiv{ \EeffMM( \rqm, \rmm ) }
    \right\rangle_{ \tilde{E} }
    .
\label{Yang_grad_alt}
\EE
This form is found to be equivalent
with the present gradient expressions,
e.g., \Eq{grad_cont}.

%-----------------------------------------------------------------------------

\section{\label{sec:mixed_rep}Mixed representation}

As seen from \Eqs{Yang_Einternal}{Yang_E_model3},
the QM/MM-MFEP method is based on a ``mixed''
representation of the QM/MM electrostatic interactions.
That is, the QM wavefunction is calculated
with the continuous \Schroedinger
equation in \Eq{SE_cont}, while the internal QM energy etc are
defined in terms of ESP charges.
In this mixed representation, $\AMF( \Rvec )$  may be defined as
\BEA
\label{AMF_mixed}
    \AMF( \Rvec )
    & = &
    \EeffQM[ \Rvec, \espSC ]
    -
    \sum_\siteA
    Q^\SC_\siteA( \Rvec ) \espSC( \Rvec_\siteA | \Rvec )
    \nonumber
    \\
    & & +
    \AMM( \Rvec, \QvecSC )
    ,
\EEA
and the mixed form of the self-consistency condition is
\BSUB
\label{SC_cond_mixed}
\BEA
\label{SC_cond_mixed_vmm}
    \espSC( \xvec | \Rvec )
    & = &
    \ll
    \vmm( \xvec, \RMM )
    \gg
    _{ \QvecSC }
    ,
    \\
    Q^\SC_\siteA( \Rvec )
    & = &
    \langle \PsiSC |
    \Qop_\siteA
    | \PsiSC \rangle
    ,
\EEA
\ESUB
where $\PsiSC = \Psi[ \Rvec, \espSC ]$.
The gradient of $\AMF( \Rvec )$ then becomes
\BW
\BEA
    \frac{ \del }{ \del \Rvec_\siteA }
    \AMF( \Rvec )
    & = &
    \left.
      \frac{ \EeffQM[ \Rvec, v' ] }{ \del \Rvec_\siteA }
    \right|
    _{ v' = \espSC }
    +
    \sum_\siteB
    \frac{ \del Q^\SC_\siteB( \Rvec ) }{ \del \Rvec_\siteA }
    \left\{
      \ll
        \vmm( \Rvec_\siteB, \RMM )
      \gg
      _{ \QvecSC }
      -
      \espSC( \Rvec_\siteB | \Rvec )
    \right\}
    \nonumber
    \\
    & & +
    Q^\SC_\siteA( \Rvec )
    \left\{
      \ll
        \frac{ \del \vmm( \Rvec_\siteA, \RMM ) }{ \del \Rvec_\siteA }
      \gg
      _{ \QvecSC }
      -
      \left.
        \frac{ \del \espSC( \Rvec_\siteA | \Xvec ) }{ \del \Rvec_\siteA }
      \right|
      _{ \Xvec = \Rvec }
    \right\}
    +
    \ll
      \frac{ \del \EeffMM( \Rvec, \RMM ) }{ \del \Rvec_\siteA }
    \gg
    _{ \QvecSC }
    \nonumber
    \\
    & & +
    \int d\xvec
      \rhoSC( \xvec | \Rvec )
      \frac{ \del \espSC( \xvec | \Rvec ) }{ \del \Rvec_\siteA }
    -
    \sum_\siteB
    Q^\SC_\siteB( \Rvec )
      \left[
        \frac{ \del \espSC( \xvec | \Rvec ) }{ \del \Rvec_\siteA }
      \right]
      _{ \xvec = \Rvec_\siteB }
    ,
\label{mixed_deriv1}
\EEA
\EW
where
$
  \rhoSC( \xvec | \Rvec )
  \equiv
  \langle \PsiSC |
  \rhoop( \xvec )
  | \PsiSC \rangle
$.
The curly brackets in the above equation vanish
by using \Eq{SC_cond_mixed_vmm} and its derivative with respect to $\xvec$
[see also \Eq{vmm_deriv}].
The third line also vanishes approximately
since $\QvecSC$ represent the ESP charges that correspond to $\rhoSC$.
Therefore, we obtain the following gradient:
\BE
\label{grad_mixed}
    \frac{ \del }{ \del \Rvec_\siteA }
    \AMF( \Rvec )
    \simeq
    \left.
      \frac{ \EeffQM[ \Rvec, v' ] }{ \del \Rvec_\siteA }
    \right|
    _{ v' = \espSC }
    +
    \ll
      \frac{ \del \EeffMM( \Rvec, \RMM ) }{ \del \Rvec_\siteA }
    \gg
    _{ \QvecSC }
    .
\EE

%-----------------------------------------------------------------------------

\section{\label{sec:non-var}Generalization to non-variational QM methods}

The main text assumes that the underlying QM wavefunction
is exact or calculated using QM methods
with variational nature (e.g., Hartree-Fock and DFT). This means
that the Hellmann-Feynman theorem holds and it can be used to
define partial charges via \Eq{partial_charge}.
However, this is not the case
for non-variational QM methods like the MP2 theory.
In the latter case, one needs to generalize
the definition of partial charges as follows,
\BE
\label{gen_partial_charge}
    \tilde{\Qvec}( \Rvec, \Phivec' )
    \equiv
    \frac
	{ \del \EeffQM( \Rvec, \Phivec' ) }
	{ \del \Phivec' }
    ,
\EE
since the first derivative of effective QM energy plays the
role of partial charges as described in \Sec{MF_pert}.
Accordingly, one needs to define the internal QM energy as
\BE
    \EQMtilde( \Rvec, \Phivec' )
    \equiv
    \EeffQM( \Rvec, \Phivec' )
    -
    \tilde{\Qvec}( \Rvec, \Phivec' )
    \cdot
    \Phivec'
    .
\EE
With these definitions the discussion in \Sec{MF_pert} remains valid.
However, the actual calculation of generalized partial charges
in \Eq{gen_partial_charge} may be tedious
unless some analytical algorithms are available.
Fortunately, in the MP2 method one can avoid such a calculation
by discarding higher-order terms in correlation
energy.\cite{Angyan_NLPT,Angyan_MP2} To see this,
let us denote relevant quantities at the MP2 level as
$\Eeff_\MP$ and $\QvecSCMP$ etc,
and the difference between the MP2 and HF levels
as $\Delta\Qvec = \QvecSCMP - \QvecSCHF$ etc.
Then, the mean-field free energy at the MP2 level may be written as
\BEA
    A_\MP
	& = &
	E_\MP( \PhivecSCMP )
	+
	\AMM( \QvecSCMP )
	\\
	& = &
	\Eeff_\MP( \PhivecSCMP )
	-
	\QvecSCMP \cdot \PhivecSCMP
	+
	\AMM( \QvecSCMP )
    \nonumber
    .
\EEA
By inserting $\QvecSCMP = \QvecSCHF + \Delta\Qvec$ and
$\Phivec^\SC_\MP = \Phivec^\SC_\HF + \Delta \Phivec$,
and making the first-order expansion in terms of
$\Delta\Qvec$ and $\Delta\Phivec$,
we have
\BEA
    A_\MP
	& = &
	\Eeff_\MP( \PhivecSCHF )
	- 
	\QvecSCHF \cdot \PhivecSCHF
	+
	\AMM( \QvecSCHF )
	+
	O( \Delta^2 )
    \nonumber
	\\
	& = &
	A_\HF
	+
	\Delta\Eeff_\MP
	+
	O( \Delta^2 )
    ,
\EEA
where
\BE
    \Delta\Eeff_\MP
    =
	\Eeff_\MP( \PhivecSCHF )
	- 
	\Eeff_\HF( \PhivecSCHF )
    .
\EE
Since $O( \Delta^2 )$ is of higher order in correlation energy,
\cite{Angyan_NLPT,Angyan_MP2}
it may safely be neglected at the MP2 level. The MP2 correction
for free energy is thus given by $\Delta\Eeff_\MP$, and we do not need to
calculate $\Qvec^\SC_\MP$ nor $\Phivec^\SC_\MP$ explicitly.
The $\Delta\Eeff_\MP$ can be evaluated using the standard expression
\BE
    \Delta \Eeff_\MP
    =
    \frac{ 1 }{ 4 }
    \sum_{abrs}
    \frac
    {
      |\langle ab || rs \rangle|^2
    }
    {
      \varepsilon_a + \varepsilon_b
      - \varepsilon_r - \varepsilon_s
    }
    ,
\EE
where $\{ \varepsilon_a \}$ etc are obtained with $\HQM + \Qvecop \cdot \PhivecSCHF$.

%=============================================================================
% References

%-- \clearpage

%-- \bibliography{refs}

\begin{thebibliography}{79}
\bibitem[{Not({\natexlab{a}})}]{Note_ESP_grid}
\bibinfo{note}{The ESP fitting grid was generated on fused sphere van der Waals
  surfaces with scaling factors 1.4, 1.5, $\ldots$, 2.5, corresponding to
  \texttt{vdwscl}=1.4, \texttt{vdwinc}=0.1, \texttt{layer}=12 in the
  \$\texttt{PDC} input group of GAMESS (\Ref{Gamess}).}
\bibitem[{Not({\natexlab{b}})}]{Note_FEP_error}
\bibinfo{note}{The FEP method may be more susceptible to the electrostatic
  periodicity effect. This is because the solvent is equilibrated to the
  unperturbed QM configuration and is thus slightly "off equilibrium" for
  perturbed configurations. The electrostatic shielding of the QM image charges
  by the solvent is thus less complete for perturbed configurations.}
\end{thebibliography}

%/////////////////////////////////////////////////////////////////////////////
\end{document}